\documentclass[12pt,reprint,superscriptaddress,amsmath,
amssymb,aps,prb,floatfix]{revtex4-2}

\usepackage{graphicx}   
\usepackage{dcolumn}    
\usepackage{bm}         
\usepackage[breaklinks=true,colorlinks,citecolor=blue,linkcolor=red,urlcolor=blue]{hyperref}   
\usepackage{amsmath}
\usepackage{amssymb} 
\usepackage{xcolor}
\usepackage{braket}
\usepackage{soul}
\usepackage[normalem]{ulem}
\usepackage[ruled,vlined,linesnumbered]{algorithm2e}

\newcommand{\eqlabel}[1]{Eq.~\eqref{#1}}
\newcommand{\applabel}[1]{Appendix~\ref{#1}}
\newcommand{\seclabel}[1]{Sec.~\ref{#1}}
\newcommand{\tablabel}[1]{Table~\ref{#1}}
\newcommand{\figlabel}[1]{Fig.~\ref{#1}}
\newcommand{\alglabel}[1]{Algorithm~\ref{#1}}

\begin{document}

\title{Constant Depth Digital-Analog Counterdiabatic Quantum Computing}

\author{Balaganchi A. Bhargava}
\email{b.a.bhargava@gmail.com}
\affiliation{Kipu Quantum GmbH, Greifswalderstrasse 212, 10405 Berlin, Germany}

\author{Shubham Kumar}
\affiliation{Kipu Quantum GmbH, Greifswalderstrasse 212, 10405 Berlin, Germany}

\author{Anne-Maria Visuri}
\affiliation{Kipu Quantum GmbH, Greifswalderstrasse 212, 10405 Berlin, Germany}

\author{Paolo A. Erdman}
\affiliation{Kipu Quantum GmbH, Greifswalderstrasse 212, 10405 Berlin, Germany}

\author{Enrique Solano}
\affiliation{Kipu Quantum GmbH, Greifswalderstrasse 212, 10405 Berlin, Germany}

\author{Narendra N. Hegade}
\affiliation{Kipu Quantum GmbH, Greifswalderstrasse 212, 10405 Berlin, Germany}

\date{\today}

\begin{abstract}
We introduce a digital-analog quantum computing framework that enables counterdiabatic protocols to be implemented at constant circuit depth, allowing fast and resource-efficient quantum state preparation on current quantum hardware. Counterdiabatic protocols suppress diabatic excitations in finite-time adiabatic evolution, but their practical application is limited by the nonlocal structure of the required Hamiltonians and the resource overhead of fully digital implementations. Counterdiabatic terms can be expressed as truncated expansions of nested commutators of the adiabatic Hamiltonian and its parametric derivative. Here, we show how this algebraic structure can be efficiently realized in a digital-analog setting using commutator product formulas. Using native multi-qubit analog interactions augmented by local single-qubit rotations, this approach enables higher-order counterdiabatic protocols whose implementation requires a constant number of analog blocks for any fixed truncation order, independent of system size. We demonstrate the method for two-dimensional spin models and analyze the associated approximation errors. These results show that digital-analog quantum computing enables a qualitatively new resource scaling for counterdiabatic protocols and related quantum control primitives, with direct implications for quantum simulation, optimization, and algorithmic state preparation on current quantum devices.
\end{abstract}

\maketitle

\section{Introduction}

Quantum computing has emerged as one of the most transformative technologies
of the past decade, driven by rapid progress in both hardware development~\cite{Lloyd1996, Peruzzo2014, Preskill2018, Arute2019, Bruzewicz2019, Zhong2020, Meissen2024, Andersen2025a, King2025a} and algorithm design~\cite{Deutsch1985, Farhi2014, Georgescu2014, Albash2018, Cerezo2021, georgescu2021,Bauer2023}. Contemporary noisy intermediate-scale quantum (NISQ) devices already enable the simulation of nontrivial many-body physics and chemistry, the execution of variational algorithms, the exploration of optimization problems, and early demonstrations of quantum advantage~\cite{Arute2019, Zhong2020, King2025a, Andersen2025a}. Nevertheless, a central challenge persists: achieving accurate quantum state preparation within the stringent coherence times and limited circuit depths of present-day quantum platforms.

Counterdiabatic (CD) driving, a prominent approach within the broader framework of shortcuts to adiabaticity, addresses this challenge by introducing auxiliary control terms that suppress diabatic transitions between instantaneous eigenstates during time-dependent quantum evolution~\cite{Berry2009, Adolfo2013, Torrontegui2013, Okuyama2017, Guery-Odelin2019}. These CD terms 
are designed such that, when added to the original Hamiltonian, the system follows the instantaneous eigenstates of the reference adiabatic Hamiltonian exactly, despite evolving over finite and potentially short times. In this sense, CD driving reproduces the adiabatic state evolution without requiring slow, adiabatic dynamics. Despite its conceptual appeal, the practical implementation of CD driving is severely constrained by the highly nonlocal structure of the CD Hamiltonians. 
One approach to constructing CD Hamiltonians is through nested commutator (NC) expansions of the adiabatic Hamiltonian and its parametric derivative~\cite{Claeys2019}, though 
the complexity of such expansions grows rapidly with system size and the truncation order. The resulting operators involve multiple non-local interaction terms that cannot be implemented natively on current quantum hardware. The applicability of CD protocols is thereby limited in practice, usually up to the first order, yielding terms with same degree of locality as the target Hamiltonian ~\cite{Sels2017, Hegade2021, Hegade2022b, Chandarana2023, Schindler2024, Dalal2024, Kumar2025aa, Kumar2025bb,visuri2025, Hegade2025}.

Alternatively, CD driving can be performed using multi-qubit analog interactions along with digital gates, under a paradigm called digital-analog quantum computing (DAQC)~\cite{Kumar2025aa}. DAQC leverages the fact that many hardware platforms, such as superconducting qubits and trapped ions, can natively generate long-range interactions~\cite{Lamata2018, Mikel2020, Julen2016, Gonzalez-Raya2021DigitalAnalog}. By incorporating DAQC, one can reduce the overall circuit depth. Digital layers provide flexibility and analog blocks are responsible for maintaining low error rates across the algorithm implementation.

Both digital and digital-analog implementations of CD driving, derived from NC, are typically restricted to the first order of the expansion, which provides a hardware-feasible Hamiltonian. Higher-order terms in the full NC expansion require many-body interactions beyond current platform capabilities and remain unexplored, limiting the scalability of the approach for complex adiabatic schedules.

In this work, we introduce a digital-analog approach to CD quantum computation that enables the implementation of the NC ansatz~\cite{Claeys2019} to arbitrary order. Our method exploits the native analog interactions available on current quantum hardware~\cite{Andersen2025a, Fors2024, Gong2021_2D_Superconducting_Array, Solomons2025_SemiGlobalTrappedIons, Shapira2025_ProgrammableGlobalDrive} and supplements them by single-qubit digital gates. We demonstrate that CD Hamiltonians derived from the NC expansion can be simulated using Trotter-Lie product formulas (PF) for commutators~\cite{Childs2013product, Childs2021} at constant circuit depth: For a fixed order of the NCs, the required circuit depth is independent of the system size. Furthermore, we propose novel decompositions of the CD terms, which, when combined with this digital-analog strategy, lead to a substantial reduction in computational resources compared to a purely digital approach. 

To illustrate the practicality and versatility of our approach, we discuss two representative case studies: the Ising spin glass model on a two-dimensional lattice and the two-dimensional XXZ spin model.
These models are of broad relevance across multiple research domains. The spin glass Hamiltonian provides a canonical mapping between spin systems and combinatorial optimization problems, with direct applications in areas such as protein folding~\cite{Alejandro2008, Romero2025, Chandarana2023}, portfolio optimization~\cite{Hegade2022b}, and supply chain logistics~\cite{Dalal2024}. The XXZ model is a foundational system for investigating correlations and collective phenomena in quantum materials described by interacting spins~\cite{sachdev2011quantum}.
These examples are implemented using distinct native interactions that are readily available on current processors such as the superconducting and trapped-ion quantum platforms~\cite{Gong2021_2D_Superconducting_Array, Hudomal2024_Floquet_XXZ_Google,Andersen2025a,Britton2012_2D_Ising_TrappedIons, Kumar2025aa}. 
Beyond these specific demonstrations, our framework provides a general and scalable strategy for simulating higher-order NC Hamiltonians at constant depth, thereby paving the way toward the realization of CD quantum computing on near-term devices.

\begin{figure*}[tb]
\centering
\includegraphics[width=1\textwidth]{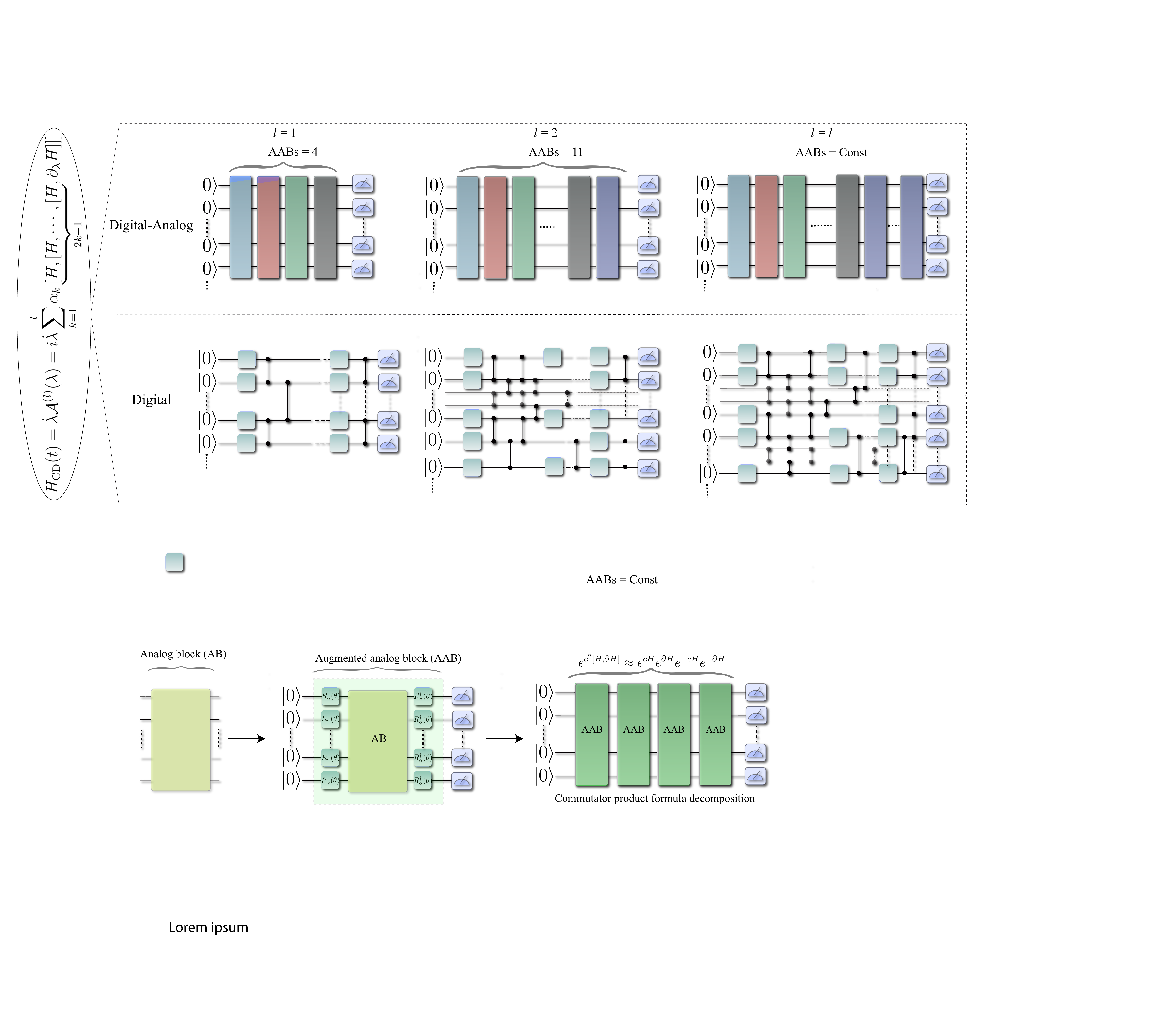}
\caption{Digital-analog counterdiabatic quantum computing (DACQC) circuit construction.
A native multi-qubit interaction implementing the hardware connectivity graph $S$ is used to define an analog block (AB), corresponding to continuous-time evolution under the native Hamiltonian.
By applying appropriate single-qubit rotations $R_{\alpha}(\theta)$, where $\alpha \in \{x, y, z\}$, before and after the interaction, one obtains augmented analog blocks (AABs) that implement the adiabatic Hamiltonian $H$ and its parametric derivative $\partial H$.
Sequences of these AABs are then composed according to product-formula constructions to generate the commutators appearing in the CD Hamiltonian.
Applying product formula decompositions recursively allows the implementation of higher-order NCs.
For illustration, the figure shows 
the first-order ($l=1$) product formula for a single time step $m$.
Repeating this construction for $M$ time steps realizes the full time evolution.}
\label{fig:1}
\end{figure*}

\section{Counterdiabatic Quantum Computing}
\label{sec:counterdiabatic_computing}

Adiabatic quantum computing~\cite{Albash2018, Barends2016aa} (AQC) is a model of computation rooted in the quantum adiabatic theorem, in which the solution to a computational problem is encoded in the ground state of a Hamiltonian. The computation begins with an initial Hamiltonian $H_{0}$, whose ground state is easily prepared, and proceeds by smoothly interpolating to a problem Hamiltonian $H_{1}$, whose ground state encodes the solution. A general way to implement this interpolation is by defining the adiabatic Hamiltonian
\begin{equation}
\label{eq:Ad-ham-def}
H(t) = (1 - \lambda(t)) H_{0} + \lambda(t) H_{1},
\end{equation}
where $\lambda(t)$ is a scheduling function that varies smoothly from $0$ to $1$ over a total evolution time $T$. If the evolution is sufficiently slow compared to the inverse square of the minimum spectral gap, the adiabatic theorem guarantees that the system remains in its instantaneous ground state, yielding the correct solution upon measurement.

On NISQ-era devices, however, long runtimes are impractical due to short coherence times. Counterdiabatic driving, a shortcut-to-adiabaticity technique~\cite{Berry2009, Hegade2021, Claeys2019, Demirplak2003, Guery-Odelin2019, Hidetoshi2021, Kiely2020, Okuyama2017, Schindler2024, Sels2017}, addresses this limitation by introducing an additional control term, $H_{\mathrm{CD}}(t)$, which suppresses diabatic transitions. The resulting modified Hamiltonian,
\begin{equation}
    \mathcal{H}(t) = H(t) + H_{\mathrm{CD}}(t),
\end{equation}
is constructed such that excitations induced by finite-time driving are canceled. This enables state preparation over significantly shorter evolution times~\cite{Hidetoshi2021} while maintaining high accuracy~\cite{Hegade2021}.

A systematic strategy for constructing $H_{\mathrm{CD}}(t)$ is provided by the NC series expansion~\cite{Claeys2019}. It is defined through sums over the NCs of the adiabatic Hamiltonian $H$ and its derivative $\partial_{\lambda}H$. We introduce the recursive notation
\begin{equation}
\mathcal{C}^{(0)} = \partial_\lambda H, 
\qquad
\mathcal{C}^{(n)} = [H, \mathcal{C}^{(n-1)}], \quad n \ge 1,
\end{equation}
so that $\mathcal{C}^{(n)}$ is an $n$-fold NC of $H$ with $\partial_\lambda H$. Using this notation, the $l$-th-order NC approximation to the CD Hamiltonian is
\begin{equation}
\label{eq:NC-AGP-def}
\begin{split}
H_{\mathrm{CD}}^{(l)}(t) 
= \dot{\lambda}(t)\,\mathcal{A}^{(l)}(\lambda(t)) = i\,\dot{\lambda}(t) \sum_{k=1}^{l} \alpha_k(t) \, \mathcal{C}^{(2k-1)} ,
\end{split}
\end{equation}
where $\mathcal{A}$ is known as the adiabatic gauge potential (AGP). The coefficients $\alpha_k$ are obtained by minimizing the action $\mathcal{S} = \mathrm{Tr}\{ G^\dagger G \}$, where  
\begin{equation}
    G = \partial_\lambda H - i[H, \mathcal{A}^{(l)}].
\end{equation} 
We omit explicit time arguments for readability. 
In the limit $l \rightarrow \infty$, Eq.~\eqref{eq:NC-AGP-def} converges to the exact CD Hamiltonian, yielding fully transitionless driving. 
CD protocols may be realized using either digital~\cite{Hegade2021} (gate-based) or analog~\cite{Qi2024} (Hamiltonian-engineered) quantum hardware, each with its own strengths and limitations.

On gate-based quantum devices, the continuous time evolution generated by the Hamiltonian $\mathcal{H}(t)$ cannot be implemented directly and must instead be approximated by a finite sequence of discrete unitary operations. The basic building blocks of this approximation are the device-native interactions $h_{\mathrm{ntv},j}$ acting on one or more qubits $j$, each of which generates a short-time unitary evolution $U_{\mathrm{ntv},j} = e^{-i \delta t \, h_{\mathrm{ntv},j}}$.
By concatenating such native unitaries over small time steps and combining them with arbitrary single-qubit rotations, one can approximate the desired many-body evolution. Together, these operations constitute a universal gate set for quantum computation~\cite{Deutsch1985}.

A standard approach to digital time evolution discretizes the time-ordered evolution operator,
\begin{equation}
\label{eq:evolution-op-def}
\mathcal{U}_{\mathcal{H}}
= \mathcal{T}\exp\left( -i\int_{0}^{T} \mathcal{H}(t)~ dt \right)
\approx \prod_{m=1}^{M} \mathcal{U}_{m},
\end{equation}
where $\mathcal{U}_{m} = e^{-i\delta t \mathcal{H}_m}$ and $\delta t = T/M$. Here, $\mathcal{T}$ denotes time ordering, and we set $\hbar = 1$.
In a digital decomposition, each short-time propagator can be implemented as $\mathcal{U}_{m} = \prod_{j} \widetilde{\Lambda}_{m,j}^\dagger \, U_{\mathrm{ntv},j} \, \widetilde{\Lambda}_{m,j}$, where $j$ is the qubit index and the $\widetilde{\Lambda}_{m,j}$ are single- or multi-qubit rotations chosen to reconstruct the desired effective Hamiltonian at step~$m$.
Such digital implementations of CD algorithms are known as \emph{digitized counterdiabatic quantum computing} (DCQC)~\cite{Hegade2021}.

Analog quantum platforms, by contrast, implement dynamics through continuous-time evolution under native interactions \(H_{\mathrm{ntv}}\), without requiring Trotterized gate decompositions. While the available controls are typically less flexible than universal digital gate sets, key Hamiltonian parameters can be tuned continuously in time. In platforms such as trapped ions, ultracold atoms, and superconducting circuits, external controls, including laser amplitudes, magnetic-field gradients, or microwave drives, are shaped to directly engineer the target Hamiltonian. As a result, the evolution \(H(t)\) is realized as a continuous control process at the Hamiltonian level rather than as a sequence of discrete circuit operations. Such approaches are particularly well suited for quantum simulation, where time-dependent control of Hamiltonian parameters enables faithful realizations of interacting many-body dynamics, as demonstrated experimentally in optical lattices and trapped-ion platforms~\cite{Bloch2012, Blatt2012, Qi2024, Lukin2024}.

For CD protocols, the auxiliary Hamiltonian $H_{\mathrm{CD}}$ can, in favorable cases, be implemented as a time-dependent continuous control field on analog platforms. When accessible, such an analog realization avoids gate-based synthesis and allows the CD term to be applied with high temporal resolution. However, the applicability of this approach is fundamentally limited by the restricted set of interactions that can be directly engineered on a given platform. In general, exact CD Hamiltonians contain nonlocal and higher-body operators that lie outside the native interaction set, necessitating truncations or approximate constructions that are strongly model- and hardware-dependent. As a result, purely analog CD protocols lack a systematic path to generalization across arbitrary Hamiltonians. Experimental demonstrations in optical-lattice and trapped-ion platforms have shown that tailored CD fields can effectively suppress diabatic excitations~\cite{Qi2024, Lukin2024, Hayasaka2023}, but these implementations rely on carefully engineered, platform-specific controls and do not readily extend to generic many-body systems.

\section{Digital-analog counterdiabatic quantum computing}

Digital-analog counterdiabatic quantum computing (DACQC) realizes CD protocols by combining a device's native multi-qubit analog interactions with a shallow layer of digital control. The central idea is to use local unitary conjugations to dress a hardware-native evolution into an effective evolution generated by a desired Hamiltonian \(\mathcal{H}(t)\). For a short time step \(\delta t\), we target
\begin{equation}
    U_{\mathcal{H}}(t_m) \equiv e^{-i \delta t\, \mathcal{H}_m}
    \approx \Lambda_m^{\dagger}\, U_{\mathrm{ntv}}(\boldsymbol{\theta}_m)\, \Lambda_m,
\end{equation}
where \(U_{\mathrm{ntv}}(\boldsymbol{\theta}_m)=e^{-i\delta t\,H_{\mathrm{ntv}}^{S}(\boldsymbol{\theta}_m)}\) is the native analog evolution available on the hardware at step \(m\), and \(\Lambda_m\) is typically a tensor product of single-qubit rotations. Because \(U_{\mathrm{ntv}}\) acts collectively on many qubits, DACQC can realize effective multi-qubit generators with substantially reduced depth compared to purely digital gate synthesis, within the constraints of the native interaction family.

Consider a hardware platform characterized by a qubit connectivity graph \(S\) and a native interaction Hamiltonian \(H_{\mathrm{ntv}}^{S}(\boldsymbol{\theta})\). We define an \emph{analog block} (AB) at step \(m\) as
\begin{equation}
    U_{\mathrm{AB},m} \equiv e^{-i \delta t\, H_{\mathrm{ntv}}^{S}(\boldsymbol{\theta}_m)},
\end{equation}
where \(\boldsymbol{\theta}_m\) denotes the programmable control parameters available on the platform (e.g., coupling strengths, detunings, or drive amplitudes). Depending on the architecture, these parameters may be tunable only within hardware constraints and may not be independently addressable on every edge of~\(S\).

To incorporate single-qubit control, we define an \emph{augmented analog block} (AAB) as
\begin{equation}
\label{eq:AAB-def}
    U_{\mathrm{AAB},m}
    \equiv \Lambda_{m}^{\dagger}\,
      U_{\mathrm{AB},m}\,
      \Lambda_{m},
\end{equation}
with \(\Lambda_m=\bigotimes_j R_j(\boldsymbol{\phi}_{j,m})\) a product of local rotations. In DACQC, the rotations \(\Lambda_m\) (and, when available, the analog parameters \(\boldsymbol{\theta}_m\)) are chosen so that \(U_{\mathrm{AAB},m}\) reproduces the desired CD generator at step~\(m\).

In the following sections, we first show how commutator PFs can be used to synthesize NC CD terms to a desired expansion order. We then construct explicit choices of $\Lambda_m$ using only single-qubit rotations for two paradigmatic native Hamiltonians relevant to current quantum architectures. Related digital-analog block constructions have been shown to improve state-preparation fidelities in prior work~\cite{Kumar2025aa, Kumar2025bb}.

\subsection{DACQC using commutator product formulas}
\label{sec:dacqc_trotter}

Figure~\ref{fig:1} shows a schematic illustration of how native interactions can be combined with Trotter PFs for CD quantum computing.  
To understand how DACQC reproduces CD dynamics, recall that the NC definition~\eqref{eq:NC-AGP-def} depends only on $H$ and $\partial_{\lambda} H$ through the set $\{\mathcal{C}^{(k)}\}$.  

Using a first-order Trotter-Suzuki decomposition~\cite{Trotter1959, Suzuki1976, Suzuki1990, Lloyd1996}, the short-time evolution operator at step $m$ is
\begin{equation}
\label{eq:trotter-split}
\begin{split}
\mathcal{U}_{m} 
&= e^{-i\delta t\big(H^{}_{m} + H_{\text{CD},m}^{(l)}\big)} \\
&= 
e^{-i\delta t\, H^{}_{m}} \,
e^{-i\delta t\, H_{\text{CD},m}^{(l)}}
       + \mathcal{O}(\delta t^2),
\end{split}
\end{equation}
where $H_{\text{CD},m}^{(l)} = \dot{\lambda}_m \mathcal{A}_{m}^{(l)} = i \dot{\lambda}_m \sum_{k=1}^{l} \alpha_{k,m} \mathcal{C}_{m}^{(2k-1)}$.
This expression contains NCs of increasing order. We can approximate each term
\[
\exp\!\left[-i\delta t\, \dot{\lambda}_{m}\, i \alpha_{k,m}\, \mathcal{C}_m^{(2k-1)}\right]
\]
by using, recursively, the group commutator identity~\cite{Childs2013product}
\begin{align}
\label{eq:childs_product_formula}
e^{A x} e^{B x} e^{-A x} e^{-B x}
= e^{[A, B] x^2 + \mathcal{O}(x^3)} .
\end{align}
This is an approximation of the exponential of a single commutator with a small parameter $x^2$, with an error that scales as
$\mathcal{O}(x^3)$.
The leading-order error scaling for NCs can be obtained by simple power counting.

For $l=1$, we define the short-time unitary
\begin{equation}
\label{eq:1nc-evo-integrated}
U^{(1)}(\delta t,\alpha_1)
= e^{-i\delta t\,H_{\text{CD}}^{(1)}} = e^{-i\delta t\,\dot{\lambda}\, i\alpha_1 \mathcal{C}^{(1)}},
\end{equation}
where $\mathcal{C}^{(1)} = [H, \partial_\lambda H]$.
Using Eq.~\eqref{eq:childs_product_formula}, and setting ${\alpha_1 > 0}$ for definiteness, we decompose this as
\begin{equation}
\label{eq:1nc-gc-def-integrated}
U^{(1)}
=
e^{-ixH}\,e^{ix\,\partial_\lambda H}\,
e^{ixH}\,e^{-ix\,\partial_\lambda H}
+ \mathcal{O}(x^3),
\end{equation}
with $x = \sqrt{|\delta t\,\dot{\lambda}\,\alpha_1|}$.
The case $\alpha_1 < 0$ is obtained by exchanging
$H \leftrightarrow \partial_{\lambda} H$.
Since the small parameter is $x^2 \propto \delta t$, the error scales as
$\mathcal{O}(\delta t^{3/2})$.

For the case $l=2$, the short-time NC evolution operator contains two commutator terms.
A first-order Trotter expansion gives an operator product of the form
\begin{align}
\label{eq:l2-trotter-split}
e^{-i\delta t\,H_{\text{CD}}^{(2)}}
= U^{(1)}(\delta t, \alpha_1)\,
U^{(3)}(\delta t, \alpha_2)
+ \mathcal{O}(\delta t^2),
\end{align}
where $U^{(3)} = e^{-i\delta t\,\dot{\lambda} i\alpha_{2}\mathcal{C}^{(3)}}$. 
To decompose $U^{(3)}$, we define
$x_{2} = \sqrt{|\delta t\,\dot{\lambda}\,\alpha_{2}|}$.
Using Eq.~\eqref{eq:childs_product_formula} recursively, we obtain
\begin{align}
\label{eq:2nc-product-formula}
&U^{(3)}(\delta t, \alpha_{2})
\nonumber \\
&= e^{-i x_2 H} e^{i x_2\mathcal{C}^{(2)}} e^{i x_2 H} e^{-i x_2\mathcal{C}^{(2)}} + \mathcal{O}(x_2^{3}) \\
\begin{split}
\label{eq:nested_gc_decomposition}
&= e^{-i x_{2} H}
\left(
e^{i \sqrt{x_{2}} H}
e^{\sqrt{x_{2}} \mathcal{C}^{(1)}}
e^{-i \sqrt{x_{2}} H}
e^{-\sqrt{x_{2}} \mathcal{C}^{(1)}}
\right) \\
&\quad \times
e^{i x_{2} H}
\left(
e^{\sqrt{x_{2}} \mathcal{C}^{(1)}}
e^{i \sqrt{x_{2}} H}
e^{-\sqrt{x_{2}}\mathcal{C}^{(1)}}
e^{-i \sqrt{x_{2}} H}
\right)
+ \mathcal{O}(x_2^{3/2}).
\end{split}
\end{align}
This expression applies for $\alpha_{2} > 0$.
The case $\alpha_{2} < 0$ is discussed in Appendix~\ref{sec:end-matter-2nd-nc-gc}.
At this level of approximation, the error scales as
$\mathcal{O}(\delta t^{3/4})$.
Further decomposing the
$e^{\pm \sqrt{x_{2}}\mathcal{C}^{(1)}}$ terms leads to factors such as
$e^{-i x_2^{1/4} H}$ and an error scaling
$\mathcal{O}(\delta t^{3/8})$.

For orders $l>2$, the same strategy applies.
We decompose the $(2l-1)$-fold NC unitary by inserting the same PF block at progressively smaller effective time scales.
Each time we ``descend" one commutator level, the corresponding small parameter is square-rooted.
After $(2l-1)$ such square-root reparameterizations, the smallest scale at which the basic block is used is
$\sim (\delta t)^{1/2^{2l-1}}$.
The error at that level is therefore

\begin{align}
\mathrm{Er}\{U^{(2l-1)}(\delta t,\alpha_l)\}
= \mathcal{O}\!\left(\delta t^{\nu_l}\right),
\qquad
\nu_l = \frac{3}{2^{2l-1}},
\label{eq:upper_bound_scaling}
\end{align}
which dominates the total error of the order-$l$ NC PF.
This scaling is an upper bound: additional contributions from higher levels only increase the power of $\delta t$.

Thus, all NC terms required for the NC approximation can be generated using only
(i) exponentials of $H$ and $\partial_\lambda H$, and
(ii) appropriate DACQC dressing of the native AB.
The first-order NC term requires only exponentials of $H$ and $\partial_\lambda H$, each of which can be implemented using AABs.
The construction of AABs for these operators is discussed for specific examples below.
For the example models, we also numerically extract the scaling of the operator-norm error with $\delta t$.

Finally, the number of exponentials that appear for a given order $l$ is
\begin{align}
\label{eq:exp-number-count}
\gamma_{H} = 2(2^{l}-1),
\qquad
\gamma_{\mathcal{C}^{(l-1)}} = 2^{l}.
\end{align}
Here, $\gamma_{H}$ denotes the number of $e^{xH}$ terms, and
$\gamma_{\mathcal{C}^{(l)}}$ denotes the number of $e^{x\,\mathcal{C}^{(l)}}$ terms.
These quantities are useful for resource estimation and for determining the number of AABs required to construct the NC of a given order $l$.

\section{Two-dimensional spin glass}
\label{sec:Ising-model}

To illustrate the power of the DACQC technique using PFs, we first consider the preparation of the ground state of the Ising spin glass model on a two-dimensional square lattice. While we focus on this geometry for concreteness, the same approach applies to any bipartite lattice with nearest-neighbor couplings in arbitrary spatial dimension. 
The model is defined as
\begin{equation}
    \label{eq:spin-glass-NN-ham}
    H_{\text{SG}} = \sum_{\langle a,b\rangle}J_{a,b}Z_{a}Z_{b} + \sum_{a = 1}^N h_{a}Z_{a},
\end{equation}
where $N$ is the number of qubits. We denote the Pauli operators acting on qubit $a$ by $X_{a}, Y_{a}$ and $Z_{a}$, and the notation $\langle a,b\rangle$ denotes pairs of nearest neighbors. 
The couplings $J_{a,b}$ and fields $h_{a}$ are drawn from uniform distributions $[-J,J)$ and $[-h, h)$, respectively. 

Starting from $H_X = -\sum_{a}X_{a}$, we interpolate to the target Hamiltonian via the parameterized Hamiltonian
\begin{equation}
    \label{eq:spin-glass-time-dependent}
    H(t) = (1-\lambda) H_X + \lambda H_{\text{SG}}.
\end{equation}
The ground state of the initial Hamiltonian~$H_X$ is $|\psi_{i} \rangle = |+\rangle^{\otimes N}$, where $|+\rangle = (|0\rangle + |1\rangle)/\sqrt{2}$.
In order to quantify the performance of the algorithms, we compute the target-state fidelity 
\begin{equation}
\label{eq:fidelity-def}
    \mathcal{F}(t) = |\langle\Psi_{f}|\psi(t)\rangle|^{2},
\end{equation}
where the reference state $|\Psi_{f}\rangle$ is the ground state of the final Hamiltonian, here \eqlabel{eq:spin-glass-NN-ham}.

The first-order NC approximation to the AGP is
\begin{equation}
\label{eq:A1_spin_glass}
    \mathcal{A}^{(1)} = -2\alpha_1 \sum_{a,b} J_{a,b}(Y_{a}Z_{b} + Z_{a}Y_{b}) - 2\alpha_1 \sum_{a}h_{a}Y_{a}, 
\end{equation} 
where $\alpha_1$ is obtained numerically as discussed in \seclabel{sec:counterdiabatic_computing}. 
Equation~\eqref{eq:A1_spin_glass} contains only two-local interactions and is therefore straightforward to implement using DACQC, provided we can construct the corresponding AABs.

\subsection{Product formula decomposition: error scaling and performance}
\label{sec:error_scaling_ising}

To verify the error scaling of the PF expansions, we compute the error in the unitary time evolution operator for a two-dimensional nearest-neighbor Ising model with random coefficients. The error, shown in \figlabel{fig:error-scaling-ising}, is defined as the operator norm $\lVert U_{\text{exact}} - U_{PF} \lVert/\sqrt{2^N}$ of the difference between the exact exponential $U_{\text{exact}}^{(l)}(\delta t) = e^{-\delta t C^{(l)}}$ and its PF expansions $U_{\text{PF}}$. The NC $C^{(1)}$ and $C^{(3)}$ appear in the first- and second-order approximations of the AGP. The operator $U_{\text{exact}}^{(1)}$ is decomposed by applying \eqlabel{eq:1nc-gc-def-integrated}, while for $U_{\text{exact}}^{(3)}$, we apply both \eqlabel{eq:2nc-product-formula} and the nested decomposition of \eqlabel{eq:nested_gc_decomposition}. To extract the scaling with $\delta t$, we use a constant value of the time-dependent factors $\dot{\lambda} = \alpha_l = 1$ in Eqs.~\eqref{eq:1nc-gc-def-integrated} and~\eqref{eq:2nc-product-formula}. The figure therefore shows only the scaling, while the magnitude of the error is different for realistic $\dot{\lambda}$ and $\alpha_l$. We find the exponent $\nu \approx 3/2$ when using \eqlabel{eq:1nc-gc-def-integrated} and \eqlabel{eq:2nc-product-formula}, as expected. For the nested PF expansion of \eqlabel{eq:nested_gc_decomposition} applied to $U_{\text{exact}}^{(3)}$, we find the exponent $\nu \approx 1.03$, which is larger than the predicted lower bound $\nu = 3/4$. In this case, therefore, the approximation is more accurate than in a general case. This is due to the fact that the first two exponential factors in \eqlabel{eq:2nc-product-formula} commute and do not contribute to the error.

\begin{figure}[tb]
    \centering
    \includegraphics[width=\columnwidth]{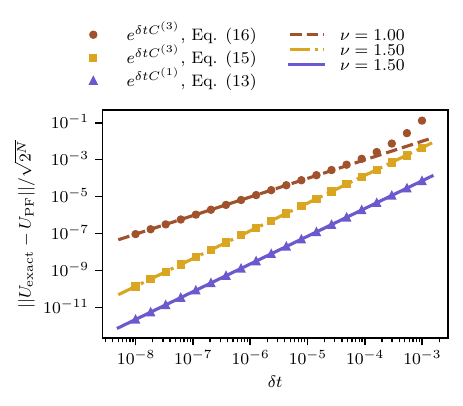}
    \caption{The errors in the PF decompositions $U_{\text{GC}}$ of the unitary operators $U_{\text{exact}}^{(l)}(\delta t) = e^{\delta t C^{(l)}}$ as functions of the time step $\delta t$. The error is defined in terms of the 2-norm $\lVert. \lVert$. 
    We fit the functions $f(\delta t) = b \, \delta t^{\nu}$ to the error, shown as solid and dashed lines, and extract the fitted exponents $\nu$. Here, we consider the two-dimensional Ising model of size $3\times 3$ with random coefficients and set $\lambda=0.5$ in \eqlabel{eq:spin-glass-time-dependent}. The exponents $\nu$ are consistent with the upper-bound scaling (lower bounds on $\nu$) presented in \seclabel{sec:dacqc_trotter}: $\nu_1 = 3/2$ for Eqs.~\eqref{eq:1nc-gc-def-integrated} and~\eqref{eq:2nc-product-formula} and $\nu_2 = 3/4$ for \eqlabel{eq:nested_gc_decomposition}.}
    \label{fig:error-scaling-ising}
\end{figure}

We further investigate the performance of the PF by computing the time-dependent target-state fidelity for a two-dimensional Ising spin glass model. 
We compare the NC expansions of order $l=1$ and $l=2$, using different PF decompositions to approximate the unitary time evolution operator. In the case $l=1$, we apply \eqlabel{eq:1nc-gc-def-integrated}, whereas for $l=2$, we apply the nested decomposition \eqlabel{eq:nested_gc_decomposition}. 

\begin{figure}[ht]
    \centering
    \includegraphics[width=\columnwidth]{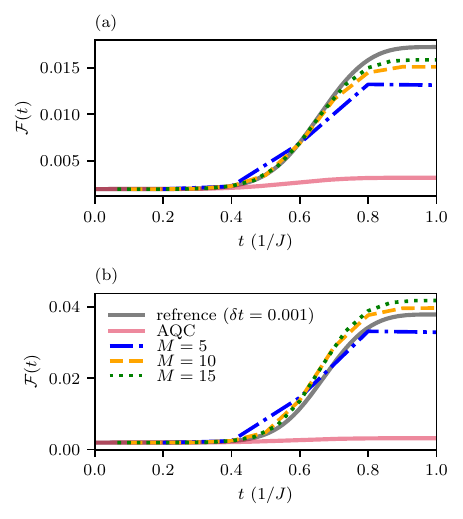}
    \caption{Target-state fidelity \eqlabel{eq:fidelity-def} as a function of instantaneous time for the Ising spin-glass of \eqlabel{eq:spin-glass-NN-ham} defined on a $3\times3$ square lattice. (a) We use \eqlabel{eq:1nc-gc-def-integrated} for the first-order NC unitary ($l=1$) and (b) \eqlabel{eq:nested_gc_decomposition} for the second-order one ($l=2$). The different colors indicate different numbers of Trotter steps $M$. The black reference line is obtained by direct exponentiation of the Hamiltonian with a small time step for which the line is converged. The couplings are random, $J_{a,b}\in (-J, J]$ and $h_{a}\in (-h, h]$, with $J=h=1$.
    }
    \label{fig:gc-fidelities-Ising}
\end{figure}

Figure~\ref{fig:gc-fidelities-Ising} shows that, as the number of Trotter steps increases-corresponding to a smaller step size, the fidelities approach the reference solution. The reference is obtained by evolving the state using \eqlabel{eq:evolution-op-def} with a sufficiently small time step for which the fidelities have converged with respect to the step size. Employing the second-order NC approximation yields higher target-state fidelities than the first-order approximation. 

When implemented as a quantum circuit, this improvement in fidelity comes at the cost of increased circuit depth with respect to the order $l$. Here, for the purpose of isolating the errors introduced by the PF, we perform the time evolution by directly exponentiating the Hamiltonian matrices at each time step. The construction of digital-analog quantum circuits for CD evolution is discussed in \seclabel{sec:daqc_implementation_spin_glass}.

Note that, depending on the native Hamiltonian of the quantum hardware, one may find simpler decompositions of $U^{(1)}$ such that the use of \eqlabel{eq:1nc-gc-def-integrated} is not necessary. These more efficient decompositions in terms of analog blocks are discussed in \seclabel{sec:daqc_implementation_spin_glass}. For this reason, in \figlabel{fig:gc-fidelities-Ising}(b) we employ \eqlabel{eq:nested_gc_decomposition} without further decomposing the operators $e^{\pm \sqrt{x_{2}}\,\mathcal{C}^{(1)}}$. We have verified that explicitly decomposing these terms yields similar results.

\subsection{DACQC: Implementing product formulas using augmented analog blocks}
\label{sec:daqc_implementation_spin_glass}

To implement the DACQC circuit, we describe how to construct the AAB for the evolution generated by the Hamiltonian $\mathcal{H}^{(1)} = H + H_{\text{CD}}^{(1)}$.
We assume that the hardware supports a native AB 
\begin{equation}
    U_{\mathrm{ntv}}^{S} = \exp\left( -i\sum_{\langle a,b\rangle}J_{a,b}Z_{a}Z_{b}\right).
\end{equation}
Analog blocks of this type are common in ion trap quantum hardware~\cite{Shapira2025_ProgrammableGlobalDrive} as well as superconducting qubits with flux-tunable couplers \cite{Fors2024}.
The two-body interaction in $H(t)$ can be implemented using a single AB for each time step $m$ by making use of the native interactions. 
The one-body terms in the Hamiltonian are implemented using single-qubit rotations.

The CD unitary $U^{(1)}$ can be implemented with two AABs, denoted by $V^{YZ}$ and $V^{ZY}$. 
To obtain the AABs, we define the following two sublattice-selective rotations on the square lattice: 
\begin{equation}
    \Lambda^{YZ} = \prod_{a\in A}R^{x}_{a}(-\pi/2) \qquad \Lambda^{ZY} = \prod_{b\in B}R^{x}_{b}(-\pi/2),
\end{equation}
where $A$ and $B$ denote the two sublattices. 
The resulting circuit synthesis procedure is outlined in \alglabel{alg:1nc-cd-zz-TFIM-one-step}.

\begin{algorithm}[t]
\caption{Two-layer synthesis of the first-order NC-CD $ZY\mid YZ$ evolution using $ZZ$ analog blocks}
\label{alg:1nc-cd-zz-TFIM-one-step}

\KwIn{Couplings $\{J_{a,b}\}$ with $a\in A$, $b\in B$; time step $\delta t$}
\KwOut{Output unitary $V_{ZY\mid YZ}(\delta t)$}

\[
\begin{aligned}
V_{ZY\mid YZ}(\delta t)
&=
\exp\!\Bigl[-i\tfrac{\delta t}{2}\sum_{a,b}J_{a,b}Y_a Z_b\Bigr] \\
&\quad \times
\exp\!\Bigl[-i\tfrac{\delta t}{2}\sum_{a,b}J_{a,b}Z_a Y_b\Bigr]
\end{aligned}
\]

\BlankLine

\textbf{Define native analog block:}
\[
U_{ZZ}(t) \gets
\exp\!\Bigl[-i\,t\sum_{a\in A,b\in B}J_{a,b}Z_a Z_b\Bigr]
\]

\BlankLine

\textbf{Define collective rotations:}
\[
\Lambda^{YZ} \gets \prod_{a\in A} R_x^{(a)}(-\pi/2),
\qquad
\Lambda^{ZY} \gets \prod_{b\in B} R_x^{(b)}(-\pi/2)
\]

\BlankLine

\textbf{Layer 1: generate $Z_a Y_b$ terms}
\[
V^{(1)} \gets
\Lambda^{ZY}\,
U_{ZZ}(\delta t/2)\,
(\Lambda^{ZY})^\dagger
\]

\BlankLine

\textbf{Layer 2: generate $Y_a Z_b$ terms}
\[
V^{(2)} \gets
\Lambda^{YZ}\,
U_{ZZ}(\delta t/2)\,
(\Lambda^{YZ})^\dagger
\]

\BlankLine

\textbf{Total unitary:}
\[
V_{ZY\mid YZ}(\delta t) \gets V^{(2)} V^{(1)}
\]

\Return{$V_{ZY\mid YZ}(\delta t)$}
\end{algorithm}

As shown in the algorithm, the operator $e^{-i\delta t ~ \mathcal{C}^{(1)}} \propto V_{ZY|YZ}(\delta t)$ can be realized using two AABs independently of the system size.
Consequently, each time step only requires a single AB implementing the two-body $ZZ$ interactions in $H(t)$ and two AABs implementing the two-body $ZY|YZ$ CD terms.
The Hamiltonian~\eqref{eq:spin-glass-NN-ham} can also be engineered using alternative native ABs such as $XX+YY$ -type interactions demonstrated in recent experiments \cite{Andersen2025a}. The details of this alternative construction are provided in \applabel{appx:Algorithms-Analaog-blocks}. 
In addition, in~\applabel{appx:digital-circ-performance}, we include a comparison between the fidelities obtained from fully digital circuits and those from the corresponding DACQC circuits.

The evolution operator corresponding to the second-order NC Hamiltonian, $\mathcal{H}^{(2)} = H + H_{\text{CD}}^{(2)}$ can now be constructed using the PF in Eq.~\eqref{eq:nested_gc_decomposition}. The resulting circuit depth is $17$, which can be straightforwardly determined by combining the PFs, the AAB decompositions, and the exponential-counting rule given in \eqlabel{eq:exp-number-count}.

\subsection{Circuit depth analysis}
\label{sec:circuit_depth_ising}

The DACQC decomposition, as described in \seclabel{sec:daqc_implementation_spin_glass}, provides a substantial circuit-depth advantage over a purely digital implementation. To enable a meaningful and robust depth comparison, we derive a lower bound $P_{\mathrm{min}}$ on the circuit depth required on a digital platform with the same qubit connectivity. The actual circuit depth in a fully digital implementation is expected to exceed this bound, as our estimate accounts for only a subset of the terms that can be executed in parallel on digital hardware. For simplicity, we restrict our analysis to a two-dimensional square-lattice connectivity.

While hardware platforms with higher connectivity could, in principle, reduce the depth of digital circuits, the implementation of higher-order NC terms inevitably leads to interactions of increasing range, which in turn require a large number of two-qubit entangling gates to decompose them. To be precise, a single $k$-body term requires $2(k-1)$ CNOT gates. In contrast, the product-formula-based DACQC approach naturally generates these higher-order interactions through native analog evolution. Digital-analog quantum simulation of higher-order Hamiltonians has been proposed in recent work~\cite{katz2025hybrid}. 

Table~\ref{tab:number-terms-Ising} lists the number of terms appearing in the Hamiltonian~\eqref{eq:spin-glass-NN-ham} as a function of the NC order $l$ and the system’s linear size $L$. To estimate a lower bound on the circuit depth required for a digital implementation, we consider only the fastest-scaling terms at each order. These contributions are highlighted in bold in the table.

As an illustrative example, consider the three-body terms at order $l=2$ in a given time step $m$. On a two-dimensional square lattice, there are $\mathcal{O}(L^{2}/3)$ three-body terms that can be implemented in parallel. Even under the optimistic assumption that each such term can be realized using a single entangling gate, the total number of layers required to implement the $18L^{2} - 12L + 12$ three-body terms is at least $54$ layers, already for a modest system size of $3 \times 3$. We emphasize that this assumption is made solely to estimate the maximum degree of parallelism and hence obtain a lower bound on the circuit depth.

By contrast, the number of AABs required to implement the full second-order NC Hamiltonian is only $14 + 1 + 2 = 17$. Implementing the full evolution operator at each time step additionally requires single-qubit gates; however, their contribution to the overall circuit depth is negligible. Consequently, in our depth estimates, we count only the number of AABs.

Clearly, this is a drastic improvement over the digital implementation. 
The number of AABs required and the approximate lower bound are given in \tablabel{tab:number-terms-Ising} for other terms as well. 
\begin{table}[tb]
\centering
\begin{tabular}{|c|c|c|c|c|}
\hline
Order & Term & No. of terms &  $P_{\mathrm{min}}$ & No. of AABs \\
\hline
$l=0$ & 1-body & $2L^2$ & - & 1 \\
      & 2-body & $2L(L-1)$ & 4 &  \\
\hline
$l=1$ & 1-body & $3L^{2}$  & - & 3 \\
      & 2-body & $\boldsymbol{6L(L-1)}$ & 12 & \\
\hline
 $l=2$ & 1-body & $3L^2$  & - & 17 \\
       & 2-body & $10L(L-1)$ &  20 & \\
       & 3-body & $\boldsymbol{6(3L(L-2) + 2)}$  & 54 & \\
       & 4-body & $4(L-1)(L-2)$ & 16 & \\
\hline
$l=3$  & 1-body & $3L^2$ & - & 79\\
       & 2-body & $10L(L-1)$ & 20 & \\
       & 3-body & $20(2+3L(L-2))$ & 120 & \\
       & 4-body & $\boldsymbol{4(L-2)(31L-27)}$ & 496 & \\
       & 5-body & $124+L(41L-148)$ & 205 & \\
\hline
\end{tabular}
\caption{Scaling information for the two-dimensional Ising spin glass. The table shows the number of terms in the Hamiltonian~\eqref{eq:spin-glass-NN-ham} on a two-dimensional grid of size $N = L\times L$. $P_{\mathrm{min}}$~is a lower bound on the number of layers required to implement the fastest-scaling terms.
The number of AABs counts the augmented analog blocks of the $ZZ$ type required to realize the corresponding Hamiltonian $\mathcal{H}^{(l)}=H + H_{\text{CD}}^{(l)}$.}
\label{tab:number-terms-Ising}  
\end{table}

\subsection{Improvement in ground-state preparation fidelity}

As discussed in the previous section, DACQC provides a substantial circuit-depth reduction over fully digital implementations. To demonstrate how this advantage translates into realistic hardware performance, we analyze the fidelity of ground-state preparation for the Hamiltonian of~\eqlabel{eq:spin-glass-NN-ham}. In particular, we compute the target-state fidelity at the final time $t = T$ using the first-order NC CD Hamiltonian. As shown in~\tablabel{tab:number-terms-Ising}, the $l=1$ implementation requires approximately four times fewer circuit layers than its purely digital counterpart. For a fixed hardware coherence time, this depth reduction enables DACQC to execute roughly four times more layers, allowing for longer effective evolution times or finer Trotter discretizations.

\begin{figure}
    \centering
    \includegraphics[width=\columnwidth]{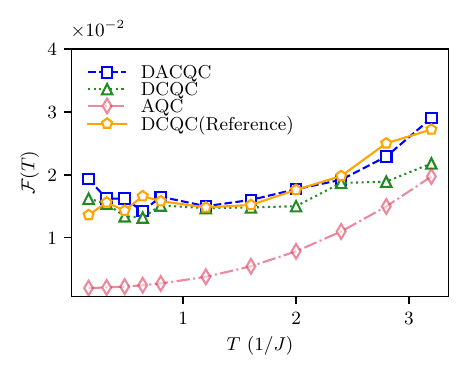}
    \caption{Target-state fidelity~\eqlabel{eq:fidelity-def} for the spin-glass Hamiltonian~\eqlabel{eq:spin-glass-NN-ham} on a $3\times 3$ lattice. 
As a benchmark, we include the fidelity obtained by evolving the initial state using exact exponentiation in adiabatic quantum computing (AQC). 
We also show the corresponding DCQC-DCQC(Reference)-results using the same time step $\delta t$ as in the DACQC simulation. 
The time step is set to $\delta t = 0.02/J$ for DACQC and $\delta t = 0.08/J$ for DCQC. 
All the three results are obtained using a circuit sampler with $20{,}000$ shots.
}
    \label{fig:TFIS-tot-time-fidelity}
\end{figure}

Figure~\ref{fig:TFIS-tot-time-fidelity} illustrates the resulting improvement in the ground-state fidelity when the DACQC circuit is implemented with a time-step size that is four times smaller than that used in the corresponding digital (DCQC) circuit. The observed increase in fidelity highlights how DACQC can be leveraged to improve state-preparation performance under constraints on circuit-depth.

\section{XXZ model}
In this section, as a further example, we discuss the ground-state preparation of the XXZ spin Hamiltonian
\begin{equation}
    H_{XXZ}= J \sum_{\langle a,b \rangle} \left(X_{a}X_{b} + Y_{a}Y_{b} + \Delta Z_{a}Z_{b} \right),
    \label{eq:xxz_hamiltonian}
\end{equation}
where $J$ is the nearest-neighbor coupling and $\Delta$ is the unitless anisotropy parameter. This model is often found as a low-energy effective description of strongly-correlated electron systems~\cite{giamarchi2003,auerbach2012interacting}. It is used as a benchmark model for quantum magnetism in compounds where spin–orbit coupling or crystal-field effects lead to anisotropic exchange interactions~\cite{Winter2017}. 

Throughout this section, we employ analog blocks of the $XX+YY$ form, which are common in superconducting qubit architectures~\cite{Arute2019, Barends2019}. 
The analog evolution with the Hamiltonian $J \sum_{\langle a,b \rangle} (X_{a}X_{b} + Y_{a}Y_{b})$, coupled with digital state preparation, was recently demonstrated in critical regimes that are challenging for classical methods~\cite{Andersen2025a}.

As the initial Hamiltonian, we consider 
\begin{equation}
\label{eq:xxz-initial-ham}
    H_Z = -\sum_{a\in A} Z_a   + \sum_{b\in B}Z_{b},
\end{equation} 
whose ground state has an antiferromagnetic spin configuration. We consider a bipartite lattice with sublattices $A$ and $B$ and the corresponding sublattice indices $a$ and $b$, respectively. Hence, the discussion applies to but is not restricted to the two-dimensional square lattice. We consider the time-dependent Hamiltonian 
\begin{equation}
\label{eq:time-dependent-xxz-hamiltonian}
H(\lambda) = (1-\lambda)H_Z +\lambda H_{XXZ}.
\end{equation}
The first-order NC approximation of the AGP can be found using Eq.~\eqref{eq:NC-AGP-def} as 
\begin{equation}
    \mathcal{A}^{(1)}(\lambda) = 4J \alpha_1(\lambda) \sum_{\langle a, b\rangle} \left( Y_{a}X_{b} - X_{a}Y_{b} \right).
\end{equation}

\subsection{Product formula decomposition}

As in \seclabel{sec:error_scaling_ising}, we analyze the errors introduced by the PF decomposition by computing the error in the operator norm $\lVert U_{\text{exact}} - U_{\text{PF}} \lVert/\sqrt{2^N}$ as a function of the time step $\delta t$. The error is shown for different PF decompositions in \figlabel{fig:error_scaling_xxz}, corresponding to \figlabel{fig:error-scaling-ising}. As for the Ising model, we find the expected lower-bound exponent $\nu = 3/2$ when using the decompositions of \eqlabel{eq:1nc-evo-integrated} and \eqlabel{eq:2nc-product-formula}. For \eqlabel{eq:nested_gc_decomposition}, we find $\nu \approx 1$, which is again larger than the lower bound $3/4$. For the parameters used in \figlabel{fig:error_scaling_xxz}, the error due to the nested decomposition of \eqlabel{eq:nested_gc_decomposition} starts to deviate from the small-$\delta t$ power-law scaling already at $\delta t \approx 10^{-5}$. This indicates that small time steps are necessary for a controlled error in the nested PF decomposition, although the magnitude of the error depends on the time-dependent factors $\alpha_l$ and $\dot{\lambda}$ not taken into account in \figlabel{fig:error_scaling_xxz}.

\begin{figure}[tb]
    \centering
    \includegraphics[width=\columnwidth]{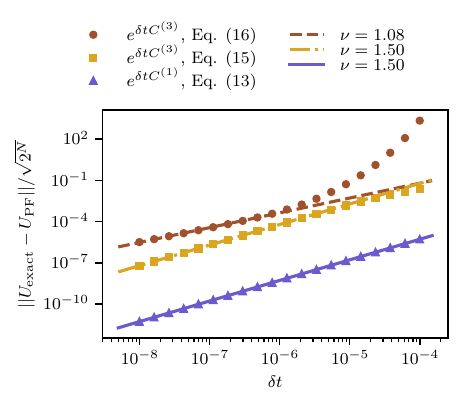}
    \caption{The errors in the PF decompositions $U_{\text{PF}}$ of the unitary operators $U_{\text{exact}}^{(l)}(\delta t) = e^{\delta t C^{(l)}}$ as functions of the time step $\delta t$, similar to \figlabel{fig:error-scaling-ising}. Here, we consider the two-dimensional XXZ model with $\Delta = 0.5$, system size $3\times 3$, and set $\lambda=0.5$ in \eqlabel{eq:time-dependent-xxz-hamiltonian}. The fitted exponents $\nu$ are consistent with the upper-bound scaling of \eqlabel{eq:upper_bound_scaling}.}
    \label{fig:error_scaling_xxz}
\end{figure}

\begin{figure}
    \centering
    \includegraphics[width=\columnwidth]{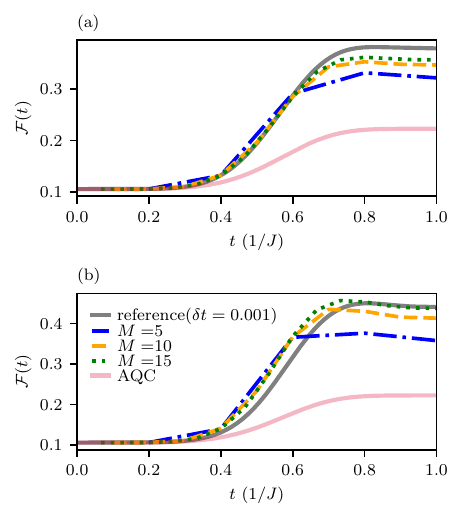}
    \caption{Target-state fidelity as a function of time for the XXZ spin model \eqlabel{eq:xxz_hamiltonian}. The CD evolution is performed with (a) first- and (b) second-order NC approximation of the AGP. In panel (a), we use the GC decomposition of \eqlabel{eq:1nc-gc-def-integrated}, and in panel (b), we apply \eqlabel{eq:nested_gc_decomposition}. For an increasing number of Trotter steps, the fidelity approaches the converged reference line. The plots are for a system of size $3\times3$ with $J = -1$ and $\Delta=0.5$. }
    \label{fig:xxz-fidelity}
\end{figure}

We further analyze the time-dependent target-state fidelity for different decompositions. Figure~\ref{fig:xxz-fidelity} shows the instantaneous fidelity with respect to the ground state of the final Hamiltonian \eqlabel{eq:xxz_hamiltonian}. The second-order NC expansion results in higher fidelities than the first-order one. Compared to the Ising spin glass model, the target-state fidelity is larger both in the initial and final state, which indicates that the chosen initial state has a larger overlap with the target state. 
The initial state is an antiferromagnetic state with a fixed total magnetization $\langle \hat{M} \rangle = \sum_{j = 1}^N \langle \hat{Z}_j \rangle$, and we verify that the ground state of the final Hamiltonian has the same total magnetization. For the systems considered here, we find that the CD evolution conserves the total magnetization with high numerical precision even though the approximate CD Hamiltonian does not commute with $\hat{M}$. We have verified that using instead the initial Hamiltonian $H_X$, as for the Ising spin glass, leads to smaller fidelities. 

\subsection{Circuit construction and depth}

As in \seclabel{sec:circuit_depth_ising}, the evolution operator associated with the Hamiltonian of \eqlabel{eq:time-dependent-xxz-hamiltonian} can be directly synthesized using the available ABs. In particular, the leading CD contribution can be generated by applying a single layer of local rotations $R_{z}^{(b)}(-\pi/2)$
acting exclusively on the $B$ sublattice. The corresponding single-layer synthesis of the CD evolution, using the first-order NC ansatz, is summarized in \alglabel{alg:CD-one-layer}.

\begin{algorithm}[t]
\caption{One-layer synthesis of the first-order NC-CD $XY\mid YX$ evolution using $(XX{+}YY)$ analog blocks}
\label{alg:CD-one-layer}

\KwIn{Couplings $\{J_{a,b}\}$ with $a\in A$, $b\in B$; time step $\delta t$}
\KwOut{Output unitary $V_{XY\mid YX}(\delta t)$}

\[
V_{XY\mid YX}(\delta t)
=
\exp\!\Bigl[-i\,\delta t \sum_{a\in A,b\in B}
4 J_{a,b}\bigl(Y_a X_b - X_a Y_b\bigr)\Bigr]
\]

\BlankLine

\textbf{Define collective rotations:}
\[
R_B \gets \prod_{b\in B} R_z^{(b)}\!\left(-\tfrac{\pi}{2}\right)
\]

\BlankLine

\textbf{Define the native analog block:}
\[
U_{XX{+}YY}(\tau) \gets
\exp\!\Bigl[-i\,\tau
\sum_{a\in A,b\in B}
4J_{a,b}\bigl(X_aX_b+Y_aY_b\bigr)\Bigr]
\]

\BlankLine

\textbf{Single-layer synthesis:}
\[
V_{XY\mid YX}(\delta t) \gets
R_B \, U_{XX{+}YY}(\delta t)\, R_B^\dagger
\]

\Return{$V_{XY\mid YX}(\delta t)$}
\end{algorithm}

Constructing the full adiabatic Hamiltonian in \eqlabel{eq:time-dependent-xxz-hamiltonian} requires additional steps. Nevertheless, all two-body interaction terms can be implemented using three AABs. In particular, the $ZZ$ interaction can be realized through a two-layer decomposition. The explicit construction is presented in Appendix~\ref{appx:Algorithms-Analaog-blocks}.
We emphasize that the above synthesis is fully general: it applies to arbitrary (including random) couplings $J_{a,b}$ and $\Delta_{a,b}$, and is valid for any bipartite lattice geometry. 

The AAB decomposition drastically reduces the circuit depth in a similar fashion as for the Ising model. 
To quantify this reduction, we follow the same depth-analysis procedure as in Sec.~\ref{sec:Ising-model}. The number of terms in the Hamiltonian appearing at different NC orders $l$ is summarized in \tablabel{tab:number-terms-XXZ}. 
\begin{table}[tb]
\centering
\begin{tabular}{|c|c|c|c|c|}
\hline
Order & Term & No. of terms &  $P_{\mathrm{min}}$ & No. of AABs \\
\hline
$l=0$ & 1-body & $L^2$ & - & 3 \\
      & 2-body & $4L(L-1)$ & 8 &  \\
\hline
$l=1$ & 1-body & $L^2$  & - & 4 \\
      & 2-body & $\boldsymbol{8L(L-1)}$ & 16 & \\
\hline
 $l=2$ & 1-body & $L^2$  & - &  26 \\
       & 2-body & $8L(L-1)$ &  16 & \\
       & 3-body & 0  & 0 & \\
       & 4-body & $\boldsymbol{4 (17 L^2 - 49L + 30)}$ & 204 & \\
\hline
$l=3$  & 1-body & $L^2$ & - &  132\\
       & 2-body & $8L(L-1)$ & 16 & \\
       & 3-body & 0 & 0 & \\
       & 4-body & $\boldsymbol{4 (330 - 415L + 119 L^2)}$ & 1904 & \\
       & 5-body & $4 (54 - 89L + 31 L^2)$ & 620 & \\
\hline
\end{tabular}
\caption{Scaling information for the two-dimensional XXZ model, similar to \tablabel{tab:number-terms-Ising}. Here, we use the AB $XX+YY$. }
\label{tab:number-terms-XXZ}
\end{table}
We have used the same assumption of the number of terms that can be implemented in parallel within each time step to estimate a lower bound on $P_{\mathrm{min}}$. 

The XXZ model can also be simulated using the $ZZ$ ABs with similar decompositions. In~\applabel{appx:digital-circ-performance}, we additionally show a comparison of the fidelities obtained from the DACQC circuit and the corresponding digital circuit for CD evolution using the first-order NC approximation.

\section{Conclusions}

We have presented an efficient approach to implement NC CD driving using digital-analog unitary operations. This is achieved by employing Lie-Trotter PF for commutators, which approximately realize the $l$-th-order CD terms at a constant circuit depth that is independent of the system size and connectivity. The advantage of this approach primarily arises from the combination of single-qubit rotations with multi-qubit gates that naturally implement different interactions. As a result, this implementation achieves a drastic reduction in circuit depth compared to purely digital realizations, making it particularly well suited for noisy intermediate-scale quantum architectures.

We demonstrate the versatility of the DACQC protocol through two paradigmatic models: the transverse-field Ising spin glass and the XXZ spin model. In each case, we derive efficient decompositions of the unitary time evolution operators into AABs, enabling the realization of the first-order NC CD terms using only one or two layers. 

Moreover, when the ABs are fully programmable, allowing for individual tuning of intra-block couplings, the same framework can be extended to effectively synthesize higher-order many-body interactions and to encode spin-glass Hamiltonians in arbitrary spatial dimensions. This capability enables applications in quantum optimization tasks, including portfolio optimization and logistics, sampling and thermal state preparation~\cite{Hegade2025}, and may also be relevant for quantum chemistry, where high-fidelity ground-state preparation is essential.

For higher-order NC terms, the approximation error of the target unitary generally grows exponentially with the order of the expansion and polynomially with the total evolution time when a fixed time-step size is used. Suppressing these errors therefore requires progressively smaller time steps, which in turn increases the number of Trotter steps in the implementation. Nevertheless, despite this exponential growth at the level of operator-norm errors, the fidelity of the prepared quantum state is expected to improve polynomially with the number of steps. This distinction arises because state-preparation fidelity is typically less sensitive to coherent unitary errors than the corresponding distance between unitaries. As a result, even approximate realizations of the CD protocol can yield substantial polynomial improvements in ground-state preparation fidelity, making DACQC particularly attractive for practical applications on near-term quantum hardware.

Despite these advantages, significant hardware challenges remain. In particular, scaling up the size and complexity of analog blocks is nontrivial, as control errors and decoherence typically increase with the complexity of the underlying hardware connectivity, leading to a degradation of effective gate fidelities. Overcoming these limitations will require advances in hardware calibration, robust pulse-level control, and error-mitigation strategies, as well as adaptive designs of ABs that balance expressivity with controllability. 
In additon, benchmarking the performance of the ABs also remains a key challenge. 
Addressing these challenges will be essential for realizing the full potential of hybrid digital-analog CD protocols on next-generation quantum processors.

\section*{Acknowledgement}
We thank Sebastián V. Romero for the inspiring discussions.

\twocolumngrid

\appendix

\section{Derivation of the second nested commutator product formula}
\label{sec:end-matter-2nd-nc-gc}

The PF for the second term $U^{(3)}$ in \eqlabel{eq:l2-trotter-split} can be derived by substituting the PF identity \eqlabel{eq:childs_product_formula} into itself.
For $ \alpha_{2} > 0$ and $x_{2} = \sqrt{|\delta t\dot{\lambda}\alpha_{2}|}$, we can write
\begin{widetext}
\begin{align*}
U^{(3)}(\delta t, \alpha_2)
&= e^{-i x_2^2 \cdot i \mathcal{C}^{(3)}} 
= e^{x_{2}^{2}  [-iH, i\mathcal{C}^{(2)}] } \\
= & e^{-ix_{2}H}e^{ix_{2}\mathcal{C}^{(2)}}e^{ix_{2}H}e^{-ix_{2}\mathcal{C}^{(2)}} + \mathcal{O}(x_2^3) \\
= & e^{-ix_{2}H}\left( e^{x_{2}[iH,\mathcal{C}^{(1)}]} \right)e^{ix_{2}H}\left( e^{x_{2}[\mathcal{C}^{(1)}, iH]} \right) + \mathcal{O}(x_2^3) \\
= & e^{-ix_{2}H}\left( e^{i\sqrt{x_{2}}H} e^{\sqrt{x_{2}}\mathcal{C}^{(1)}}e^{-i\sqrt{x_{2}}H}e^{-\sqrt{x_{2}}\mathcal{C}^{(1)}} \right)e^{ix_{2}H}\left( e^{\sqrt{x_{2}}\mathcal{C}^{(1)} }e^{i\sqrt{x_{2}}H} e^{-\sqrt{x_{2}} \mathcal{C}^{(1)}}e^{-i\sqrt{x_{2}}H}  \right) 
+ \mathcal{O}(x_2^{\frac{3}{2}}).
\end{align*}
\end{widetext}
To implement the first-order term $U^{(1)}$, one can directly use an efficient decomposition such as the AABs in~\alglabel{alg:1nc-cd-zz-TFIM-one-step} or \alglabel{alg:CD-one-layer}. Alternatively, the AABs that produce $H$ and $\partial_{\lambda}H$ can be used to implement \eqlabel{eq:1nc-gc-def-integrated}.
With the replacement $ \alpha \rightarrow \sqrt{c_{2}}$, we have
\begin{align*}
e^{\sqrt{x_{2}}\mathcal{C}^{(1)}} &= e^{\sqrt{x_{2}}[H, \partial_{\lambda} H]} = e^{\sqrt{x_{2}}[-iH, i\partial_{\lambda} H]} \\
&= e^{-ix_{2}^{1/4}H}e^{ix_{2}^{1/4}\partial_{\lambda} H} e^{ix_{2}^{1/4}H}e^{-ix_{2}^{1/4}\partial_{\lambda} H} + \mathcal{O}(x_2^{\frac{3}{4}})
\end{align*}
and $e^{-\sqrt{x_{2}}\mathcal{C}^{(1)}} = e^{\sqrt{x_{2}}[\partial_{\lambda} H, H]}$ is obtained by exchanging $H \leftrightarrow \partial_{\lambda} H$.

For the case $\alpha_{2} < 0$, we have 
\begin{align*}
U^{(3)}&(\delta t, \alpha_2) 
= e^{i x_2^2 \cdot i C^{(3)}} = e^{x_2^2 [iH, iC^{(2)}]} \\
= & e^{ix_{2}H} 
\left( e^{i\sqrt{x_{2}}H} e^{\sqrt{x_{2}}\mathcal{C}^{(1)}}
e^{-i\sqrt{x_{2}}H}
e^{-\sqrt{x_{2}}\mathcal{C}^{(1)}} \right)
e^{-ix_{2}H} \\
&\times \left( e^{\sqrt{x_{2}}\mathcal{C}^{(1)}}
e^{i\sqrt{x_{2}}H} e^{-\sqrt{x_{2}} \mathcal{C}^{(1)}}e^{-i\sqrt{x_{2}}H} \right) 
+ \mathcal{O}(x_2^{\frac{3}{2}}).
\end{align*}

\section{Algorithms for augmented analog blocks}
\label{appx:Algorithms-Analaog-blocks}

In this section, we provide all the algorithms that can be used to generate the AABs. 

\subsection{Ising spin glass}

First, we consider the example of the nearest-neighbor Ising spin glass, \eqlabel{eq:spin-glass-NN-ham}, defined on lattice $S$. 
Since we have framed the algorithm in terms of sublattices $A$ and $B$, the derivation here applies to any bipartite lattice in arbitrary spatial dimensions.
To obtain the AAB circuit using the $(XX+YY)$ kind of ABs, we first consider the two-body $ZZ$ term in Hamiltonian~\eqref{eq:spin-glass-NN-ham}. The corresponding unitary $\exp\!\Bigl(
-i\delta t
\sum_{a\in A,b\in B}
\widetilde{J}_{a,b} Z_a Z_b
\Bigr)$ can be approximated with the two-layer AAB decomposition given in \alglabel{alg:two_layer_zz}. The unitary 
\begin{equation}
\exp\!\Bigl[
  -i\delta t
  \sum_{a,b}
  J_{a,b}\,
  (Y_a Z_b + Z_a Y_b)
\Bigr]
\end{equation}
corresponding to the first-order NC terms can be approximated using a single $(XX+YY)$ AB,
as described in \alglabel{alg:CD-one-step-simplified}.

\begin{algorithm}[t]
\caption{Two-layer synthesis of $ZZ$ evolution using $(XX{+}YY)$ analog blocks}
\label{alg:two_layer_zz}

\KwIn{Couplings $\{\widetilde{J}_{a,b}=\Delta J_{a,b}\}$ for $a\in A$, $b\in B$; time step $\delta t$}
\KwOut{Output unitary $V_{ZZ}(\delta t)$}

\[
V_{ZZ}(\delta t)=
\exp\!\Bigl[-i\,\delta t
\sum_{a\in A,b\in B}\widetilde{J}_{a,b} Z_a Z_b\Bigr]
\]

\BlankLine

\textbf{Define the native analog block:}
\[
U_{XX{+}YY}(t) \gets
\exp\!\Bigl[-i\,t
\sum_{a\in A,b\in B}
\widetilde{J}_{a,b}\bigl(X_aX_b+Y_aY_b\bigr)\Bigr]
\]

\BlankLine

\textbf{Define single-qubit rotations:}
\[
R_1 \gets
\Bigl(\prod_{a\in A} R_y^{(a)}(\tfrac{\pi}{2})\Bigr)
\Bigl(\prod_{b\in B} R_y^{(b)}(\tfrac{\pi}{2})\Bigr)
\]
\[
R_2 \gets
\Bigl(\prod_{a\in A} R_x^{(a)}(\tfrac{\pi}{2})R_z^{(a)}(\tfrac{\pi}{2})\Bigr)
\Bigl(\prod_{b\in B} R_x^{(b)}(\tfrac{\pi}{2})R_z^{(b)}(-\tfrac{\pi}{2})\Bigr)
\]

\BlankLine

\textbf{Layer 1:}
\[
V^{(1)} \gets
R_1\,U_{XX{+}YY}(\delta t/2)\,R_1^\dagger
\]

\BlankLine

\textbf{Layer 2:}
\[
V^{(2)} \gets
R_2\,U_{XX{+}YY}(\delta t/2)\,R_2^\dagger
\]

\BlankLine

\textbf{Total unitary:}
\[
V_{ZZ}(\delta t) \gets V^{(1)} V^{(2)}
\]

\Return{$V_{ZZ}(\delta t)$}
\end{algorithm}

\begin{algorithm}[t]
\caption{One-layer synthesis of the first-order NC-CD $ZY\mid YZ$ evolution using $(XX{+}YY)$ AB}
\label{alg:CD-one-step-simplified}

\KwIn{Couplings $\{J_{a,b}\}$ with $a\in A$, $b\in B$; time step $\delta t$}
\KwOut{Output unitary $V_{ZY\mid YZ}(\delta t)$}

\[
V_{ZY\mid YZ}(\delta t)
=
\exp\!\Bigl[-i\,\delta t
\sum_{a\in A,b\in B}
J_{a,b}\bigl(Y_a Z_b + Z_a Y_b\bigr)\Bigr]
\]

\BlankLine

\textbf{Define collective rotations:}
\[
R_A \gets
\prod_{a\in A} R_y^{(a)}\!\left(-\tfrac{\pi}{2}\right),
\qquad
R_B \gets
\prod_{b\in B}
R_y^{(b)}\!\left(\tfrac{\pi}{2}\right)
R_z^{(b)}\!\left(\tfrac{\pi}{2}\right)
\]

\BlankLine

\textbf{Define the native analog block:}
\[
U_{XX{+}YY}(\tau) \gets
\exp\!\Bigl[-i\,\tau
\sum_{a\in A,b\in B}
J_{a,b}\bigl(X_aX_b+Y_aY_b\bigr)\Bigr]
\]

\BlankLine

\textbf{Single-layer synthesis:}
\[
V_{ZY\mid YZ}(\delta t) \gets
R_B\, R_A\, U_{XX{+}YY}(\delta t)\, R_A^\dagger R_B^\dagger
\]

\Return{$V_{ZY\mid YZ}(\delta t)$}
\end{algorithm}

\subsection{XXZ model}

To implement the unitary evolution due to the $ZZ$ terms of the XXZ Hamiltonian~\eqref{eq:xxz_hamiltonian}, we can use the same two-layer decomposition as for the Ising spin glass model, given in \alglabel{alg:two_layer_zz}. 

\section{Performance comparison}
\label{appx:digital-circ-performance}

In this section, we provide additional information on the performance of the DACQC decomposition based on AABs and fully digital circuit implementations. To this end, we perform state-vector simulations of both approaches. In \seclabel{sec:ising-performance}, we present the target-state fidelity and examine the different error sources in the two decompositions. In \seclabel{appx:perfomance_xxz}, we additionally show the scaling of the operator norm error and discuss the effect of operator ordering in the decompositions.

\subsection{Ising spin glass}
\label{sec:ising-performance}

Here, we compare the performance of the DACQC decomposition based on AABs with a digital circuit for the two-dimensional Ising spin glass. 
The resulting fidelities are benchmarked against those obtained from direct exponentiation of the Hamiltonian, both for the same number of Trotter steps and in the continuous-time limit~$\delta t \to 0$.
In all the simulations in this work, we have chosen a smooth scheduling function 
\begin{equation}
    \lambda(t) = \sin^2\!\left(\frac{\pi}{2}\, \sin^2\!\left(\frac{\pi t}{2T}\right)\right).
\end{equation}

Specifically, we evaluate the fidelity for preparing the ground state of the two-dimensional Ising model defined in \eqlabel{eq:spin-glass-NN-ham}, using the first-order NC construction implemented via \alglabel{alg:1nc-cd-zz-TFIM-one-step}. In this case, it is not necessary to use PF decompositions. The initial state $|+\rangle^{\otimes N}$ is evolved in time under the effective Hamiltonian $\mathcal{H}^{(1)}(t)$. The target-state fidelity as a function of instantaneous time is shown in \figlabel{fig:EM:TFISG-fidelity}.

\begin{figure}[tb!]
    \centering
    \includegraphics[width=\columnwidth]{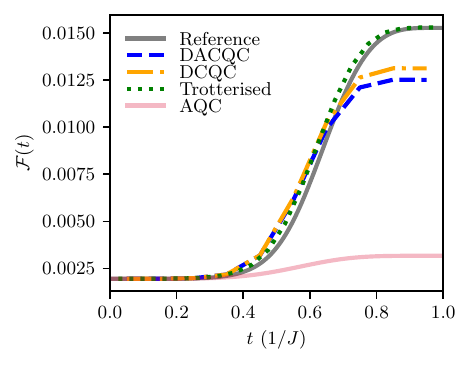}
    \caption{Time-dependent target-state fidelity [\eqlabel{eq:fidelity-def}] for the spin glass Hamiltonian~\eqref{eq:spin-glass-NN-ham}. The system size is $3\times3$ and the couplings are random, $J_{a,b}\in (-J, J]$ and $h_{a}\in (-h, h]$, with $J=h=1$.
    For these parameters and $M=10$ time steps, the fidelity obtained from the digital circuit exceeds that of the DACQC circuit.
    For the reference curve, we use $\delta t=0.001$, for which the curve is converged in $\delta t$.}
    \label{fig:EM:TFISG-fidelity}
\end{figure}

We observe that the fidelities obtained from both circuit implementations differ slightly from those produced by exact exponentiation of the Hamiltonian. 
This discrepancy arises from the Trotter decomposition of $e^{-i \delta t \mathcal{H}_m^{(1)}}$ within each time step~$m$. The first-order Trotter decompositions
\begin{align}
\label{eq:ising_trotter}
    e^{-i \delta t \mathcal{H}_m} 
    &= e^{-i \delta t H_m} e^{-i \delta t H^{(1)}_{\mathrm{CD}},m} + \mathcal{O}(\delta t^2), \\
    e^{-i \delta t H_m} 
    &= e^{-i \delta t (1-\lambda_m) H_X} e^{-i \delta t \lambda_m H_{\text{SG}}} + \mathcal{O}(\delta t^2)
\end{align}
give rise to errors in the operator norm of $e^{-i \delta t \mathcal{H}_m}$. These errors scale quadratically with the time step size and linearly with the number of non-commuting terms that must be implemented sequentially, proportional to the system size, $\mathcal{O}(N \delta t^2)$. This error is the same for the digital and DACQC decompositions, and the leading contributions stem from non-commuting terms involving $X$, $ZZ$, $Y$, and mixed operators of the form $ZY$ and $YZ$.

In a fully digital circuit, one additionally decomposes each unitary in \eqlabel{eq:ising_trotter} into one- and two-qubit gates, which for the $U^{(1)}_m = e^{-i \delta t H^{(1)}_{\mathrm{CD}},m}$ term has the same error scaling $\mathcal{O}(N \delta t^2)$. 
In contrast, in DACQC, this second Trotter decomposition is absent. Instead, the two-layer AAB decomposition of $e^{-i \delta t \sum ZZ}$ incurs an error that scales as $\mathcal{O}(N \delta t^2)$ and does not appear in the digital decomposition. Meanwhile, the single-layer AAB representation of $U^{(1)}_m$ is exact in the ideal case.

We find that these different error sources in the DCQC and DACQC circuits contribute a small difference in the target-state fidelity in \figlabel{fig:EM:TFISG-fidelity} for the system size considered here. 

The exact exponentiation of the Hamiltonian does not suffer from such decomposition-induced errors and is closest to the reference values. It would be interesting to see whether the larger number of non-commuting terms in higher-order NC expansions would lead to a larger difference between the DCQC and DACQC results.

\subsection{XXZ model}
\label{appx:perfomance_xxz}

We further compare the errors and performance of fully digital and DACQC circuits for the XXZ model. To analyze the error in the operator norm of the two decompositions, we compute the 2-norm of the difference between the exact and approximate unitary time evolution operators as a function of $\delta t$. The error of each decomposition is shown in~\figlabel{fig:XXZ-scaling-digital-aab}, similar to Figs.~\ref{fig:error-scaling-ising} and~\ref{fig:error_scaling_xxz} of the main text.
The error corresponding to the adiabatic unitary $e^{-i \delta t H_m}$ is shown at $\lambda=0.5$ in ~\figlabel{fig:XXZ-scaling-digital-aab}(a).
The errors here arise from the non-commutation of $XX + YY$ with $ZZ$ and $Z$. 
In the digital decomposition, further errors, scaling as $\mathcal{O}(N \delta t^2)$, arise from decomposing $e^{-i \delta t J \sum_{\langle a, b \rangle} (X_a X_b + Y_a Y_b)}$ into two-qubit gates. Meanwhile, the two-layer AAB decomposition of $e^{-i \delta t \sum ZZ}$ incurs an error that also scales as $\mathcal{O}(N \delta t^2)$. Despite the different error sources, we find that the errors from the two decompositions coincide. They show the expected scaling $\mathcal{O}(\delta t^2)$.

\begin{figure}[h]
    \centering
    \includegraphics[width=0.5\textwidth]{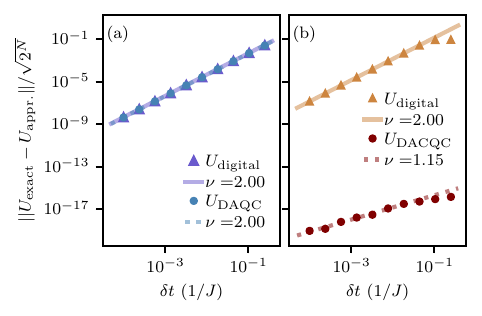}
    \caption{(a) The operator norm errors of the digital and DAQC decompositions of $U_{\text{exact}} = e^{-i \delta t H(\lambda=0.5)}$ as functions of the time step size $\delta t$. The error is defined as the scaled 2-norm $\lVert. \lVert$. The errors of the two decompositions overlap. (b) The operator norm errors of the digital and DACQC decompositions of $U_{\text{exact}} = e^{\delta t C^{(1)}}$. 
    The fits $f(\delta t) = b \, \delta t^{\nu}$ are shown as solid and dotted lines, and the fitted exponents are denoted by $\nu$. Here, we consider the two-dimensional XXZ model of size $3\times 3$, $\Delta-0.5$, and $\lambda=0.5$ in \eqlabel{eq:spin-glass-time-dependent}.
    }
    \label{fig:XXZ-scaling-digital-aab}
\end{figure}

In~\figlabel{fig:XXZ-scaling-digital-aab}(b), we plot separately the operator norm error of the first-order NC unitary $e^{\delta t C^{(1)}}$. The errors in the digital decomposition arise from the commutators of $XY$ and $YX$, and we find the expected scaling $\mathcal{O}(\delta t^2)$. The single-AAB representation of the NC unitary is exact in the ideal case. Numerically, we find an error that is smaller than the digital one by several orders of magnitude.

We also compute the time-dependent target-state fidelity as in \seclabel{sec:ising-performance}, using the full unitary time evolution operator. The fidelities are shown in \figlabel{fig:XXZ-fidelity}. 

We find that the fidelities obtained from the digital and DACQC circuits are very close to each other for the parameters used in \figlabel{fig:XXZ-fidelity}, indicating that the error from \eqlabel{eq:ising_trotter} is more significant here than the errors resulting from further decomposing the factors $e^{-i \delta t H_m}$ and $e^{-i \delta t H^{(1)}_{\text{CD}, m}}$.

\begin{figure}[h]
    \centering
    \includegraphics[width=0.5\textwidth]{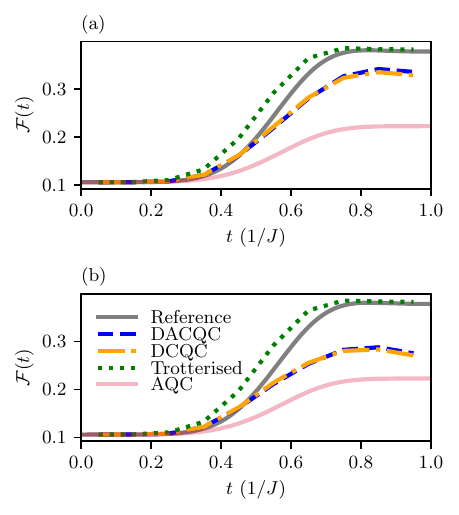}
    \caption{The target-state fidelity as a function of time for the XXZ Hamiltonian of \eqlabel{eq:xxz_hamiltonian}, defined on a square lattice of size $3 \times 3$. Here, we use the first-order NC approximation and set the parameters as $J = -1$ and $\Delta = 0.5$. We use $M=10$ Trotter steps. Since we compute the fidelities in the middle of each time step at $t_m = (m - 0.5) \delta t$, the dashed and dotted lines end at $t = 0.95/J$. (a) Each time step applies the sequence $e^{-i \delta t H}\, e^{-i \delta t H_{\mathrm{CD}}} \ket{\psi}$, while in panel (b) the order is reversed, $e^{-i \delta t H_{\mathrm{CD}}}\, e^{-i \delta t H} \ket{\psi}$.}
    \label{fig:XXZ-fidelity}
\end{figure}

For a small number of Trotter steps, such as $M = 10$ shown in~\figlabel{fig:XXZ-fidelity}, the ordering in which the different terms are applied has a noticeable impact on the resulting fidelity. To illustrate this effect, we compare two circuit constructions in which the positions of the CD and adiabatic unitaries are exchanged. In panel (a), each time step applies the sequence $e^{-i \delta t H}\, e^{-i \delta t H_{\mathrm{CD}}} \ket{\psi}$,
whereas in panel (b) the order is reversed, $e^{-i \delta t H_{\mathrm{CD}}}\,  e^{-i \delta t H}\ket{\psi}$.
Within the circuit implementing the adiabatic evolution $e^{-i \delta t H}$, the terms are arranged as
\begin{equation*}
e^{-i \delta t J \sum_{\langle a, b \rangle} (X_a X_b + Y_a Y_b)}\,
e^{-i \delta t \left( \Delta J \sum_{\langle a, b \rangle} Z_a Z_b + H_Z \right)} \ket{\psi}.
\end{equation*}
The ordering used in panel (b) yields fidelities that are slightly closer to the reference. As the step size is reduced to $\delta t \lesssim 0.01/J$, all fidelity curves converge toward the reference result, becoming essentially independent of the operator ordering.

\newpage 

\bibliographystyle{apsrev4-2}
\bibliography{references} 

@incollection{Torrontegui2013,
	author = {Erik Torrontegui and Sara Ib{\'a}{\~n}ez and Sofia Mart{\'\i}nez-Garaot and Michele Modugno and Adolfo {del Campo} and David Gu{\'e}ry-Odelin and Andreas Ruschhaupt and Xi Chen and Juan Gonzalo Muga},
	booktitle = {Advances in Atomic, Molecular, and Optical Physics},
	doi = {https://doi.org/10.1016/B978-0-12-408090-4.00002-5},
	editor = {Ennio Arimondo and Paul R. Berman and Chun C. Lin},
	issn = {1049-250X},
	pages = {117-169},
	publisher = {Academic Press},
	series = {Advances In Atomic, Molecular, and Optical Physics},
	title = {Chapter 2 - Shortcuts to Adiabaticity},
	url = {https://www.sciencedirect.com/science/article/pii/B9780124080904000025},
	volume = {62},
	year = {2013}}

@article{Adolfo2013,
	author = {del Campo, Adolfo},
	doi = {10.1103/PhysRevLett.111.100502},
	issue = {10},
	journal = {Phys. Rev. Lett.},
	month = {Sep},
	numpages = {5},
	pages = {100502},
	publisher = {American Physical Society},
	title = {Shortcuts to Adiabaticity by Counterdiabatic Driving},
	url = {https://link.aps.org/doi/10.1103/PhysRevLett.111.100502},
	volume = {111},
	year = {2013}}

@misc{Farhi2014,
	author = {Farhi, Edward and Goldstone, Jeffrey and Gutmann, Sam},
	copyright = {arXiv.org perpetual, non-exclusive license},
	doi = {10.48550/ARXIV.1411.4028},
	publisher = {arXiv},
	title = {A Quantum Approximate Optimization Algorithm},
	url = {https://arxiv.org/abs/1411.4028},
	year = {2014}}

@misc{Hayasaka2023,
	author = {Hayasaka, Hiroshi and Imoto, Takashi and Matsuzaki, Yuichiro and Kawabata, Shiro},
	copyright = {arXiv.org perpetual, non-exclusive license},
	doi = {10.48550/ARXIV.2305.08352},
	publisher = {arXiv},
	title = {A general method to construct mean field counter diabatic driving for a ground state search},
	url = {https://arxiv.org/abs/2305.08352},
	year = {2023}}

@misc{Lukin2024,
	author = {Lukin, Alexander and Schiffer, Benjamin F. and Braverman, Boris and Cantu, Sergio H. and Huber, Florian and Bylinskii, Alexei and Amato-Grill, Jesse and Maskara, Nishad and Cain, Madelyn and Wild, Dominik S. and Samajdar, Rhine and Lukin, Mikhail D.},
	copyright = {arXiv.org perpetual, non-exclusive license},
	doi = {10.48550/ARXIV.2405.21019},
	publisher = {arXiv},
	title = {Quantum quench dynamics as a shortcut to adiabaticity},
	url = {https://arxiv.org/abs/2405.21019},
	year = {2024}}

@misc{Fors2024,
	author = {Fors, Simon Pettersson and Fern{\'a}ndez-Pend{\'a}s, Jorge and Kockum, Anton Frisk},
	copyright = {Creative Commons Attribution 4.0 International},
	doi = {10.48550/ARXIV.2408.15402},
	publisher = {arXiv},
	title = {Comprehensive explanation of ZZ coupling in superconducting qubits},
	url = {https://arxiv.org/abs/2408.15402},
	year = {2024}}

@article{Barends2019,
	author = {Barends, R. and Quintana, C. M. and Petukhov, A. G. and Chen, Yu and et. al.},
	doi = {10.1103/PhysRevLett.123.210501},
	issue = {21},
	journal = {Phys. Rev. Lett.},
	month = {Nov},
	numpages = {6},
	pages = {210501},
	publisher = {American Physical Society},
	title = {Diabatic Gates for Frequency-Tunable Superconducting Qubits},
	url = {https://link.aps.org/doi/10.1103/PhysRevLett.123.210501},
	volume = {123},
	year = {2019}}

@article{Dalal2024,
	author = {Dalal, Archismita and Montalban, Iraitz and Hegade, Narendra N. and Cadavid, Alejandro Gomez and Solano, Enrique and Awasthi, Abhishek and Vodola, Davide and Jones, Caitlin and Weiss, Horst and F\"uchsel, Gernot},
	doi = {10.1103/PhysRevApplied.22.064068},
	issue = {6},
	journal = {Phys. Rev. Appl.},
	month = {Dec},
	numpages = {14},
	pages = {064068},
	publisher = {American Physical Society},
	title = {Digitized counterdiabatic quantum algorithms for logistics scheduling},
	url = {https://link.aps.org/doi/10.1103/PhysRevApplied.22.064068},
	volume = {22},
	year = {2024}}

@article{Hegade2022b,
	author = {Hegade, N. N. and Chandarana, P. and Paul, K. and Chen, Xi and Albarr\'an-Arriagada, F. and Solano, E.},
	doi = {10.1103/PhysRevResearch.4.043204},
	issue = {4},
	journal = {Phys. Rev. Res.},
	month = {Dec},
	numpages = {9},
	pages = {043204},
	publisher = {American Physical Society},
	title = {Portfolio optimization with digitized counterdiabatic quantum algorithms},
	url = {https://link.aps.org/doi/10.1103/PhysRevResearch.4.043204},
	volume = {4},
	year = {2022}}

@article{Chandarana2023,
	author = {Chandarana, Pranav and Hegade, Narendra N. and Montalban, Iraitz and Solano, Enrique and Chen, Xi},
	doi = {10.1103/PhysRevApplied.20.014024},
	issue = {1},
	journal = {Phys. Rev. Appl.},
	month = {Jul},
	numpages = {16},
	pages = {014024},
	publisher = {American Physical Society},
	title = {Digitized Counterdiabatic Quantum Algorithm for Protein Folding},
	url = {https://link.aps.org/doi/10.1103/PhysRevApplied.20.014024},
	volume = {20},
	year = {2023}}

@article{Alejandro2008,
	author = {Perdomo, Alejandro and Truncik, Colin and Tubert-Brohman, Ivan and Rose, Geordie and Aspuru-Guzik, Al\'an},
	doi = {10.1103/PhysRevA.78.012320},
	issue = {1},
	journal = {Phys. Rev. A},
	month = {Jul},
	numpages = {15},
	pages = {012320},
	publisher = {American Physical Society},
	title = {Construction of model Hamiltonians for adiabatic quantum computation and its application to finding low-energy conformations of lattice protein models},
	url = {https://link.aps.org/doi/10.1103/PhysRevA.78.012320},
	volume = {78},
	year = {2008}}

@article{Romero2025,
	author = {Romero, S. V. and Cadavid, A. G. and Nika{\v c}evi{\'c}, P and Solano, E. and et.al.},
	journal = {arXiv:2506.07866v2},
	month = {June},
	title = {Protein folding with an all-to-all trapped-ion quantum computer},
	year = {2025}}

@article{Kumar2025bb,
	author = {Kumar, Shubham and Hegade, Narendra N and Henrique de Oliveira, Murilo and Solano, Enrique and Gomez Cadavid, Alejandro and Albarr{\'a}n-Arriagada, F},
	doi = {10.1088/2058-9565/ad8b64},
	journal = {Quantum Science and Technology},
	month = {nov},
	number = {1},
	pages = {015023},
	publisher = {IOP Publishing},
	title = {Digital-analog counterdiabatic quantum optimization with trapped ions},
	url = {https://doi.org/10.1088/2058-9565/ad8b64},
	volume = {10},
	year = {2024}}

@article{Kumar2025aa,
	author = {Kumar, Shubham and Hegade, Narendra N. and Visuri, Anne-Maria and Bhargava, Balaganchi A. and Hernandez, Juan F. R. and Solano, E. and Albarr{\'a}n-Arriagada, F. and Barrios, G. Alvarado},
	date = {2025/03/11},
	doi = {10.1038/s41534-025-01001-4},
	id = {Kumar2025},
	isbn = {2056-6387},
	journal = {npj Quantum Information},
	number = {1},
	pages = {43},
	title = {Digital-analog quantum computing of fermion-boson models in superconducting circuits},
	url = {https://doi.org/10.1038/s41534-025-01001-4},
	volume = {11},
	year = {2025}}

@article{Qi2024,
	author = {Zhang, Qi and Hegade, Narendra N. and Cadavid, Alejandro Gomez and Lassabli{\`e}re, Lucas and Trautmann, Jan and Perseguers, S{\'e}bastien and Solano, Enrique and Henriet, Lo{\"\i}c and Michon, Eric},
	journal = {arXiv:2405.14829v1},
	month = {May},
	title = {Analog Counterdiabatic Quantum Computing},
	year = {2024}}

@article{Barends2016aa,
	author = {Barends, R. and Shabani, A. and Lamata, L. and Kelly, J. and et. al.},
	date = {2016/06/01},
	doi = {10.1038/nature17658},
	id = {Barends2016},
	isbn = {1476-4687},
	journal = {Nature},
	number = {7606},
	pages = {222--226},
	title = {Digitized adiabatic quantum computing with a superconducting circuit},
	url = {https://doi.org/10.1038/nature17658},
	volume = {534},
	year = {2016}}

@article{Albash2018,
	author = {Albash, Tameem and Lidar, Daniel A.},
	doi = {10.1103/RevModPhys.90.015002},
	issue = {1},
	journal = {Rev. Mod. Phys.},
	month = {Jan},
	numpages = {64},
	pages = {015002},
	publisher = {American Physical Society},
	title = {Adiabatic quantum computation},
	url = {https://link.aps.org/doi/10.1103/RevModPhys.90.015002},
	volume = {90},
	year = {2018}}

@article{Guery-Odelin2019,
	author = {Gu{\'e}ry-Odelin, David and Ruschhaupt, Andreas and Kiely, Anthony and Torrontegui, Erik and Mart{\'\i}nez-Garaot, Sara and Muga, Juan Gonzalo},
	doi = {10.1103/RevModPhys.91.045001},
	journal = {Rev. Mod. Phys.},
	number = {4},
	pages = {045001},
	title = {Shortcuts to adiabaticity: Concepts, methods, and applications},
	volume = {91},
	year = {2019}}

@article{Berry2009,
	author = {Berry, M. V.},
	doi = {10.1088/1751-8113/42/36/365303},
	journal = {J. Phys. A: Math. Theor.},
	number = {36},
	pages = {365303},
	title = {Transitionless quantum driving},
	volume = {42},
	year = {2009}}

@article{Demirplak2003,
	author = {Demirplak, Mustafa and Rice, Stuart A},
	doi = {10.1021/jp030708a},
	journal = {J. Phys. Chem. A},
	number = {46},
	pages = {9937--9945},
	title = {Adiabatic population transfer with control fields},
	volume = {107},
	year = {2003}}

@article{Cerezo2021,
	author = {Cerezo, M. and Arrasmith, A. and Babbush, R. and Benjamin, S. C. and Endo, S. and Fujii, K. and McClean, J. R. and Mitarai, K. and Yuan, X. and Cincio, L. and Coles, P. J.},
	doi = {10.1038/s42254-021-00348-9},
	journal = {Nat. Rev. Phys.},
	pages = {625--644},
	title = {Variational quantum algorithms},
	volume = {3},
	year = {2021}}

@article{Peruzzo2014,
	author = {Peruzzo, Alberto and McClean, Jarrod and Shadbolt, Peter and Yung, Man-Hong and Zhou, Xiao-Qi and Love, Peter J and Aspuru-Guzik, Al{\'a}n and O'Brien, Jeremy L},
	doi = {10.1038/ncomms5213},
	journal = {Nat. Commun.},
	pages = {4213},
	title = {A variational eigenvalue solver on a photonic quantum processor},
	volume = {5},
	year = {2014}}

@article{Georgescu2014,
	author = {Georgescu, Iulia M and Ashhab, Sahel and Nori, Franco},
	doi = {10.1103/RevModPhys.86.153},
	journal = {Rev. Mod. Phys.},
	pages = {153--185},
	title = {Quantum simulation},
	volume = {86},
	year = {2014}}

@article{Zhong2020,
	author = {Zhong, Han-Sen and others},
	doi = {10.1126/science.abe8770},
	journal = {Science},
	number = {6523},
	pages = {1460--1463},
	title = {Quantum computational advantage using photons},
	volume = {370},
	year = {2020}}

@article{Sels2017,
	author = {Sels, Dries and Polkovnikov, Anatoli},
	doi = {10.1073/pnas.1619826114},
	journal = {Proc. Natl. Acad. Sci. USA},
	number = {20},
	pages = {E3909--E3916},
	title = {Minimizing irreversible losses in quantum systems by local counterdiabatic driving},
	volume = {114},
	year = {2017}}

@article{Childs2021,
	author = {Childs, Andrew M. and Su, Yuan and Tran, Minh C. and Wiebe, Nathan and Zhu, Shuchen},
	doi = {10.1103/PhysRevX.11.011020},
	journal = {Phys. Rev. X},
	number = {1},
	pages = {011020},
	title = {Theory of Trotter error with commutator scaling},
	volume = {11},
	year = {2021}}

@article{Bruzewicz2019,
	author = {Bruzewicz, Colin D. and Chiaverini, John and McConnell, Robert and Sage, Jeremy M.},
	doi = {10.1063/1.5088164},
	journal = {Applied Physics Reviews},
	number = {2},
	pages = {021314},
	publisher = {AIP Publishing},
	title = {Trapped-Ion Quantum Computing: Progress and Challenges},
	url = {https://doi.org/10.1063/1.5088164},
	volume = {6},
	year = {2019}}

@article{Hidetoshi2021,
	author = {Prielinger, Luise and Hartmann, Andreas and Yamashiro, Yu and Nishimura, Kohji and Lechner, Wolfgang and Nishimori, Hidetoshi},
	doi = {10.1103/PhysRevResearch.3.013227},
	issue = {1},
	journal = {Phys. Rev. Res.},
	month = {Mar},
	numpages = {13},
	pages = {013227},
	publisher = {American Physical Society},
	title = {Two-parameter counter-diabatic driving in quantum annealing},
	url = {https://link.aps.org/doi/10.1103/PhysRevResearch.3.013227},
	volume = {3},
	year = {2021}}

@article{Trotter1959,
	author = {Trotter, H. F.},
	doi = {10.1090/S0002-9939-1959-0108732-6},
	issue = {4},
	journal = {Proceedings of the American Mathematical Society},
	month = {Aug},
	numpages = {7},
	pages = {545--551},
	publisher = {American Mathematical Society},
	title = {On the product of semi-groups of operators},
	url = {https://www.ams.org/journals/proc/1959-010-04/S0002-9939-1959-0108732-6/},
	volume = {10},
	year = {1959}}

@article{Suzuki1976,
	author = {Suzuki, Masuo},
	doi = {10.1007/BF01609348},
	issue = {2},
	journal = {Communications in Mathematical Physics},
	month = {Jun},
	numpages = {8},
	pages = {183--190},
	publisher = {Springer},
	title = {Generalized Trotter's formula and systematic approximants of exponential operators and inner derivations with applications to many-body problems},
	url = {https://link.springer.com/article/10.1007/BF01609348},
	volume = {51},
	year = {1976}}

@article{Suzuki1990,
	author = {Suzuki, Masuo},
	doi = {10.1016/0375-9601(90)90962-N},
	issue = {6},
	journal = {Physics Letters A},
	month = {Jun},
	numpages = {5},
	pages = {319--323},
	publisher = {Elsevier},
	title = {Fractal decomposition of exponential operators with applications to many-body theories and Monte Carlo simulations},
	url = {https://www.sciencedirect.com/science/article/pii/037596019090962N},
	volume = {146},
	year = {1990}}

@article{Lloyd1996,
	author = {Lloyd, Seth},
	doi = {10.1126/science.273.5278.1073},
	issue = {5278},
	journal = {Science},
	month = {Aug},
	numpages = {6},
	pages = {1073--1078},
	publisher = {American Association for the Advancement of Science},
	title = {Universal quantum simulators},
	url = {https://www.science.org/doi/10.1126/science.273.5278.1073},
	volume = {273},
	year = {1996}}

@article{Childs2013product,
	author = {Childs, Andrew M. and Wiebe, Nathan},
	doi = {10.1063/1.4811386},
	issn = {0022-2488},
	journal = {Journal of Mathematical Physics},
	month = {06},
	number = {6},
	pages = {062202},
	title = {Product formulas for exponentials of commutators},
	url = {https://doi.org/10.1063/1.4811386},
	volume = {54},
	year = {2013}}

@article{georgescu2021,
	author = {Georgescu, I M and Ashhab, S and Nori, F},
	doi = {10.1103/PRXQuantum.2.017003},
	journal = {PRX Quantum},
	number = {1},
	pages = {017003},
	title = {Quantum Simulators: Architectures and Opportunities},
	volume = {2},
	year = {2021}}

@article{Bauer2023,
	author = {Bauer, Christian W. and Davoudi, Zohreh and Balantekin, A. Baha and et.al.},
	doi = {10.1103/PRXQuantum.4.027001},
	issue = {2},
	journal = {PRX Quantum},
	month = {May},
	numpages = {70},
	pages = {027001},
	publisher = {American Physical Society},
	title = {Quantum Simulation for High-Energy Physics},
	url = {https://link.aps.org/doi/10.1103/PRXQuantum.4.027001},
	volume = {4},
	year = {2023}}

@article{Meissen2024,
	author = {Miessen, Alexander and Egger, Daniel J. and Tavernelli, Ivano and Mazzola, Guglielmo},
	doi = {10.1103/PRXQuantum.5.040320},
	issue = {4},
	journal = {PRX Quantum},
	month = {Nov},
	numpages = {19},
	pages = {040320},
	publisher = {American Physical Society},
	title = {Benchmarking Digital Quantum Simulations Above Hundreds of Qubits Using Quantum Critical Dynamics},
	url = {https://link.aps.org/doi/10.1103/PRXQuantum.5.040320},
	volume = {5},
	year = {2024}}

@article{King2025a,
	author = {Andrew D. King and Alberto Nocera and Marek M. Rams and Jacek Dziarmaga and et.al.},
	doi = {10.1126/science.ado6285},
	journal = {Science},
	number = {6743},
	pages = {199-204},
	title = {Beyond-classical computation in quantum simulation},
	url = {https://www.science.org/doi/abs/10.1126/science.ado6285},
	volume = {388},
	year = {2025}}

@article{Andersen2025a,
	author = {Andersen, T. I. and Astrakhantsev, N. and Karamlou, A. H. and Berndtsson, J. and et.al.},
	date = {2025/02/01},
	doi = {10.1038/s41586-024-08460-3},
	id = {Andersen2025},
	isbn = {1476-4687},
	journal = {Nature},
	number = {8049},
	pages = {79--85},
	title = {Thermalization and criticality on an analogue--digital quantum simulator},
	url = {https://doi.org/10.1038/s41586-024-08460-3},
	volume = {638},
	year = {2025}}

@article{Arute2019,
	author = {Arute, Frank and Arya, Kunal and Babbush, Ryan and Bacon, Dave and et.al.},
	date = {2019/10/01},
	doi = {10.1038/s41586-019-1666-5},
	id = {Arute2019},
	isbn = {1476-4687},
	journal = {Nature},
	number = {7779},
	pages = {505--510},
	title = {Quantum supremacy using a programmable superconducting processor},
	url = {https://doi.org/10.1038/s41586-019-1666-5},
	volume = {574},
	year = {2019}}

@article{Okuyama2017,
	author = {Okuyama, Manaka and Takahashi, Kazuya},
	doi = {10.7566/JPSJ.86.043002},
	journal = {J. Phys. Soc. Jpn.},
	number = {4},
	pages = {043002},
	title = {Counterdiabatic Formalism of Shortcuts to Adiabaticity},
	volume = {86},
	year = {2017}}

@article{Kiely2020,
	author = {Kiely, Anthony},
	doi = {10.22331/qv-2020-10-05-45},
	journal = {{Quantum Views}},
	month = oct,
	pages = {45},
	publisher = {{Verein zur F{\"{o}}rderung des Open Access Publizierens in den Quantenwissenschaften}},
	title = {Generalised {C}ounterdiabatic {D}riving in {O}pen {S}ystems},
	url = {https://doi.org/10.22331/qv-2020-10-05-45},
	volume = {4},
	year = {2020}}

@article{Hegade2021,
	author = {Hegade, Narendra N. and Paul, Koushik and Ding, Yongcheng and Sanz, Mikel and Albarr\'an-Arriagada, F. and Solano, Enrique and Chen, Xi},
	doi = {10.1103/PhysRevApplied.15.024038},
	issue = {2},
	journal = {Phys. Rev. Appl.},
	month = {Feb},
	numpages = {13},
	pages = {024038},
	publisher = {American Physical Society},
	title = {Shortcuts to Adiabaticity in Digitized Adiabatic Quantum Computing},
	url = {https://link.aps.org/doi/10.1103/PhysRevApplied.15.024038},
	volume = {15},
	year = {2021}}

@article{Schindler2024,
	author = {Schindler, Paul M. and Bukov, Marin},
	doi = {10.1103/PhysRevLett.133.123402},
	issue = {12},
	journal = {Phys. Rev. Lett.},
	month = {Sep},
	numpages = {8},
	pages = {123402},
	publisher = {American Physical Society},
	title = {Counterdiabatic Driving for Periodically Driven Systems},
	url = {https://link.aps.org/doi/10.1103/PhysRevLett.133.123402},
	volume = {133},
	year = {2024}}

@article{Claeys2019,
	author = {Claeys, Pieter W. and Pandey, Mohit and Sels, Dries and Polkovnikov, Anatoli},
	doi = {10.1103/PhysRevLett.123.090602},
	issue = {9},
	journal = {Phys. Rev. Lett.},
	month = {Aug},
	numpages = {7},
	pages = {090602},
	publisher = {American Physical Society},
	title = {Floquet-Engineering Counterdiabatic Protocols in Quantum Many-Body Systems},
	url = {https://link.aps.org/doi/10.1103/PhysRevLett.123.090602},
	volume = {123},
	year = {2019}}

@article{Preskill2018,
	author = {Preskill, John},
	doi = {10.22331/q-2018-08-06-79},
	issn = {2521-327X},
	journal = {{Quantum}},
	month = aug,
	pages = {79},
	publisher = {{Verein zur F{\"{o}}rderung des Open Access Publizierens in den Quantenwissenschaften}},
	title = {Quantum {C}omputing in the {NISQ} era and beyond},
	url = {https://doi.org/10.22331/q-2018-08-06-79},
	volume = {2},
	year = {2018}}

@article{Deutsch1985,
	author = {Deutsch, David},
	doi = {10.1098/rspa.1985.0070},
	journal = {Proceedings of the Royal Society of London. A. Mathematical and Physical Sciences},
	number = {1818},
	pages = {97-117},
	title = {Quantum theory, the Church--Turing principle and the universal quantum computer},
	url = {https://royalsocietypublishing.org/doi/abs/10.1098/rspa.1985.0070},
	volume = {400},
	year = {1985}}

@article{Blatt2012,
	author = {Blatt, Rainer and Roos, Christian F.},
	date = {2012/04/01},
	doi = {10.1038/nphys2259},
	id = {Blatt2012},
	journal = {Nature Physics},
	number = {4},
	pages = {277--284},
	title = {Quantum simulations with trapped ions},
	url = {https://doi.org/10.1038/nphys2259},
	volume = {8},
	year = {2012}}

@article{Bloch2012,
	author = {Bloch, Immanuel and Dalibard, Jean and Nascimb{\`e}ne, Sylvain},
	date = {2012/04/01},
	doi = {10.1038/nphys2252},
	id = {Bloch2012},
	journal = {Nature Physics},
	number = {4},
	pages = {267--276},
	title = {Quantum simulations with ultracold quantum gases},
	url = {https://doi.org/10.1038/nphys2252},
	volume = {8},
	year = {2012}}

@misc{Hegade2025,
	archiveprefix = {arXiv},
	author = {Narendra N. Hegade and Nachiket L. Kortikar and Balaganchi A. Bhargava and Juan F. R. Hern{\'a}ndez and Alejandro Gomez Cadavid and Pranav Chandarana and Sebasti{\'a}n V. Romero and Shubham Kumar and Anton Simen and Anne-Maria Visuri and Enrique Solano and Paolo A. Erdman},
	eprint = {2510.26735},
	primaryclass = {quant-ph},
	title = {Digitized Counterdiabatic Quantum Sampling},
	url = {https://arxiv.org/abs/2510.26735},
	year = {2025}}

@book{sachdev2011quantum,
	author = {Sachdev, Subir},
	edition = {2nd},
	publisher = {Cambridge University Press},
	title = {Quantum Phase Transitions},
	url = {https://doi.org/10.1017/CBO9780511973765},
	year = {2011}}

@book{auerbach2012interacting,
	author = {Auerbach, Assa},
	publisher = {Springer Science \& Business Media},
	title = {Interacting electrons and quantum magnetism},
	year = {2012}}

@article{Britton2012_2D_Ising_TrappedIons,
	author = {Britton, J. W. and Sawyer, B. C. and Keith, A. C. and Wang, C.-C. J. and Freericks, J. K. and Uys, H. and Biercuk, M. J. and Bollinger, J. J.},
	doi = {10.1038/nature10981},
	journal = {Nature},
	number = {7395},
	pages = {489--492},
	title = {Engineered two-dimensional Ising interactions in a trapped-ion quantum simulator},
	url = {https://www.nature.com/articles/nature10981},
	volume = {484},
	year = {2012}}

@article{Gong2021_2D_Superconducting_Array,
	author = {Gong, M. and Chen, M.-C. and Zheng, Y. and others},
	doi = {10.1126/science.abg7812},
	journal = {Science},
	number = {6545},
	pages = {948--952},
	title = {Quantum walks on a programmable two-dimensional 62-qubit superconducting processor},
	url = {https://www.science.org/doi/10.1126/science.abg7812},
	volume = {372},
	year = {2021}}

@article{Hudomal2024_Floquet_XXZ_Google,
	author = {Hudomal, A. and Prosen, T. and others},
	doi = {10.1103/PRXQuantum.5.010316},
	journal = {PRX Quantum},
	number = {1},
	pages = {010316},
	title = {Integrability Breaking and Bound States in Google's Floquet XXZ Quantum Circuit},
	url = {https://link.aps.org/doi/10.1103/PRXQuantum.5.010316},
	volume = {5},
	year = {2024}}

@article{Lamata2018,
	author = {Lucas Lamata and Adrian Parra-Rodriguez and Mikel Sanz and Enrique Solano},
	doi = {10.1080/23746149.2018.1457981},
	journal = {Advances in Physics: X},
	number = {1},
	pages = {1457981},
	title = {Digital‐Analog Quantum Simulations with Superconducting Circuits},
	url = {https://doi.org/10.1080/23746149.2018.1457981},
	volume = {3},
	year = {2018}}

@article{Mikel2020,
	author = {A. Parra-Rodriguez and P. Lougovski and L. Lamata and E. Solano and M. Sanz},
	journal = {Phys. Rev. A},
	pages = {022305},
	title = {Digital-analog quantum computations},
	url = {https://link.aps.org/doi/10.1103/PhysRevA.101.022305},
	volume = {101},
	year = {2020}}

@article{Julen2016,
	author = {Arrazola, I{\~n}igo and Pedernales, Julen S. and Lamata, Lucas and Solano, Enrique},
	journal = {Scientific Reports},
	pages = {30534},
	title = {Digital-Analog Quantum Simulation of Spin Models in Trapped Ions},
	url = {https://doi.org/10.1038/srep30534},
	volume = {6},
	year = {2016}}

@article{Gonzalez-Raya2021DigitalAnalog,
	author = {Gonzalez-Raya, Tasio and Asensio-Perea, Rodrigo and Martin, Ana and C{\'e}leri, Lucas~C. and Sanz, Mikel and Lougovski, Pavel and Dumitrescu, Eugene~F.},
	doi = {10.1103/PRXQuantum.2.020328},
	journal = {PRX Quantum},
	number = {2},
	pages = {020328},
	publisher = {American Physical Society},
	title = {Digital-Analog Quantum Simulations Using the Cross-Resonance Effect},
	volume = {2},
	year = {2021}}

@article{Solomons2025_SemiGlobalTrappedIons,
	archiveprefix = {arXiv},
	author = {Solomons, Yakov and Kadish, Yotam and Peleg, Lee and Nemirovsky, Jonathan and Ben Kish, Amit and Shapira, Yotam},
	eprint = {2509.14331},
	journal = {arXiv preprint},
	month = {sep},
	primaryclass = {quant-ph},
	title = {Full programmable quantum computing with trapped-ions using semi-global fields},
	url = {https://arxiv.org/abs/2509.14331},
	year = {2025}}

@article{Shapira2025_ProgrammableGlobalDrive,
	author = {Shapira, Y. and Markov, J. and Akerman, N. and Stern, A. and Ozeri, R.},
	doi = {10.1103/PhysRevLett.134.010602},
	journal = {Physical Review Letters},
	number = {1},
	pages = {010602},
	title = {Programmable quantum simulations on a trapped-ion quantum computer with a global drive},
	url = {https://journals.aps.org/prl/abstract/10.1103/PhysRevLett.134.010602},
	volume = {134},
	year = {2025}}

@article{katz2025hybrid,
  title         = {Hybrid digital-analog protocols for simulating quantum multi-body interactions},
  author        = {Katz, Or and Schuckert, Alexander and Wang, Tianyi and Crane, Eleanor and Gorshkov, Alexey V. and Cetina, Marko},
  journal       = {arXiv preprint arXiv:2512.21385},
  year          = {2025},
  eprint        = {2512.21385},
  primaryClass  = {quant-ph},
  doi           = {10.48550/arXiv.2512.21385}
}

@misc{visuri2025,
      title={Digitized counterdiabatic quantum critical dynamics}, 
      author={Anne-Maria Visuri and Alejandro Gomez Cadavid and Balaganchi A. Bhargava and Sebastián V. Romero and András Grabarits and Pranav Chandarana and Enrique Solano and Adolfo del Campo and Narendra N. Hegade},
      year={2025},
      eprint={2502.15100},
      archivePrefix={arXiv},
      primaryClass={quant-ph},
      url={https://arxiv.org/abs/2502.15100}, 
}

@article{Winter2017,
doi = {10.1088/1361-648X/aa8cf5},
url = {https://doi.org/10.1088/1361-648X/aa8cf5},
year = {2017},
month = {nov},
publisher = {IOP Publishing},
volume = {29},
number = {49},
pages = {493002},
author = {Winter, Stephen M and Tsirlin, Alexander A and Daghofer, Maria and van den Brink, Jeroen and Singh, Yogesh and Gegenwart, Philipp and Valentí, Roser},
title = {Models and materials for generalized Kitaev magnetism},
journal = {Journal of Physics: Condensed Matter}
}

@book{giamarchi2003,
  title={Quantum physics in one dimension},
  author={Giamarchi, Thierry},
  volume={121},
  year={2003},
  publisher={Clarendon press}
}
\end{document}